\documentclass[sigconf]{acmart}
\graphicspath{{figures/}}
\copyrightyear{2026}
\acmYear{2026}
\setcopyright{cc}
\setcctype{by}
\acmConference[CCS '26]
{Proceedings of the 2026 ACM SIGSAC Conference on Computer and Communications Security}
{November 15--19, 2026}
{The Hague, Netherlands}
\acmBooktitle{Proceedings of the 2026 ACM SIGSAC Conference on Computer and Communications Security (CCS '26), November 15--19, 2026, The Hague, Netherlands}
\acmDOI{10.1145/3830454.3846657}
\acmISBN{979-8-4007-2871-6/2026/11}

\usepackage{subcaption}

\newcommand{\greencheck}{\textcolor{green}{\ding{51}}} %
\newcommand{\redcross}{\textcolor{red}{\ding{55}}} %
\newcommand{\name}{\texttt{PhantomTac}}

\begin{document}

\title{{\name}: Hijacking Robotic Tactile Systems with Deploy-and-forget Tags}

\author{Binbin Zhang}
\email{sjtu\_zhangbb@sjtu.edu.cn}
\affiliation{%
  \institution{Shanghai Jiao Tong University}
  \city{Shanghai}
  \country{China}
}

\author{Yifeng Zhang}
\email{yifeng\_zhang@sjtu.edu.cn}
\affiliation{%
  \institution{Shanghai Jiao Tong University}
  \city{Shanghai}
  \country{China}
}

\author{Xiaomeng Chen}
\email{sjtu\_chenxm@sjtu.edu.cn}
\affiliation{%
  \institution{Shanghai Jiao Tong University}
  \city{Shanghai}
  \country{China}
}

\author{Dongyao Chen}
\authornote{Dongyao Chen is the corresponding author.}
\email{dychen@smu.edu.sg}
\affiliation{%
  \institution{Singapore Management University}
  \country{Singapore}
}

\begin{abstract}
Tactile perception is crucial for precise robotic manipulation, providing an important foundation for dexterous, closed-loop force control. 
Among various tactile sensing techniques, magnetic tactile sensors stand out thanks to their fine-grained, three-dimensional force perception capability and strong robustness against visual occlusion. 
In this study, we discover a physical loophole called {\name} in magnetic tactile sensors. 
The attacker can alter the magnetic field, thus spoofing the force interpreted by magnetic tactile sensors.
The attacker can stealthily deploy such an attack by simply attaching a thin (the thickness is around 0.5 mm), battery-free magnetic tag to the surface of an object.
Hence, the attack tag can be easily concealed by a barcode sticker.
Such an attack scheme is ``deploy-and-forget'' --- the attacker can deploy the tag and leave.
When a robotic hand is picking up an object, it may severely damage the object.
Specifically, spoofing the normal force can crush the object, whereas spoofing the shear force can lead to a loose grip, thus dropping the object.
Compared with directly sabotaging the robotic hand, {\name} is much more concealed as the attacker does not need access to the device.
We performed extensive reverse-engineering efforts to characterize the sensing mechanism of commercial-off-the-shelf tactile sensors and proposed a practical tag design. 
End-to-end evaluations conducted on real robots and robotic arms demonstrate that our attack can bypass existing anti-interference mechanisms and effectively hijack their closed-loop control processes, achieving an attack success rate exceeding 95\% when the grasping-position offset is bounded within 8~mm.
Finally, based on the limitations of existing anti-magnetic interference mechanisms, we propose a multimodal-consistency defense mechanism to mitigate such physical-layer security threats.
\end{abstract}



\maketitle
\section{Introduction}
\label{sec:intro}
\begin{figure}[t]
  \centering
  \begin{subfigure}{0.95\linewidth}
    \centering
    \includegraphics[width=1.05\linewidth]{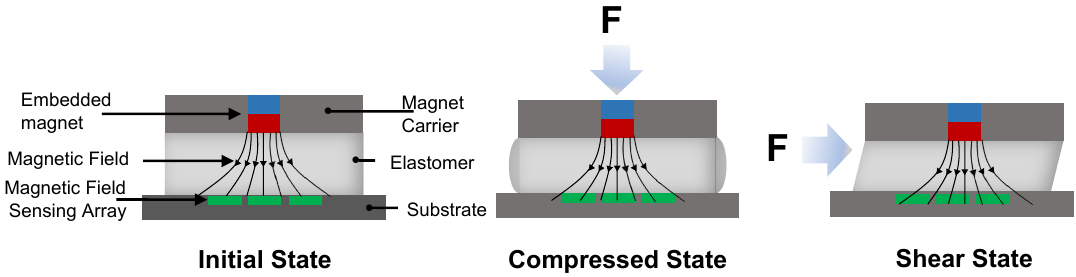}
    \Description{
    Illustration of the compressed and shear states of a tactile sensor, showing how normal and shear forces are represented during tactile interaction.
    }
    \caption{The compressed and shear states characterize normal and shear forces, respectively.}
  \end{subfigure}
  \hspace{2pt}  %
  \begin{subfigure}{0.95\linewidth}
    \centering
    \includegraphics[width=1.05\linewidth]{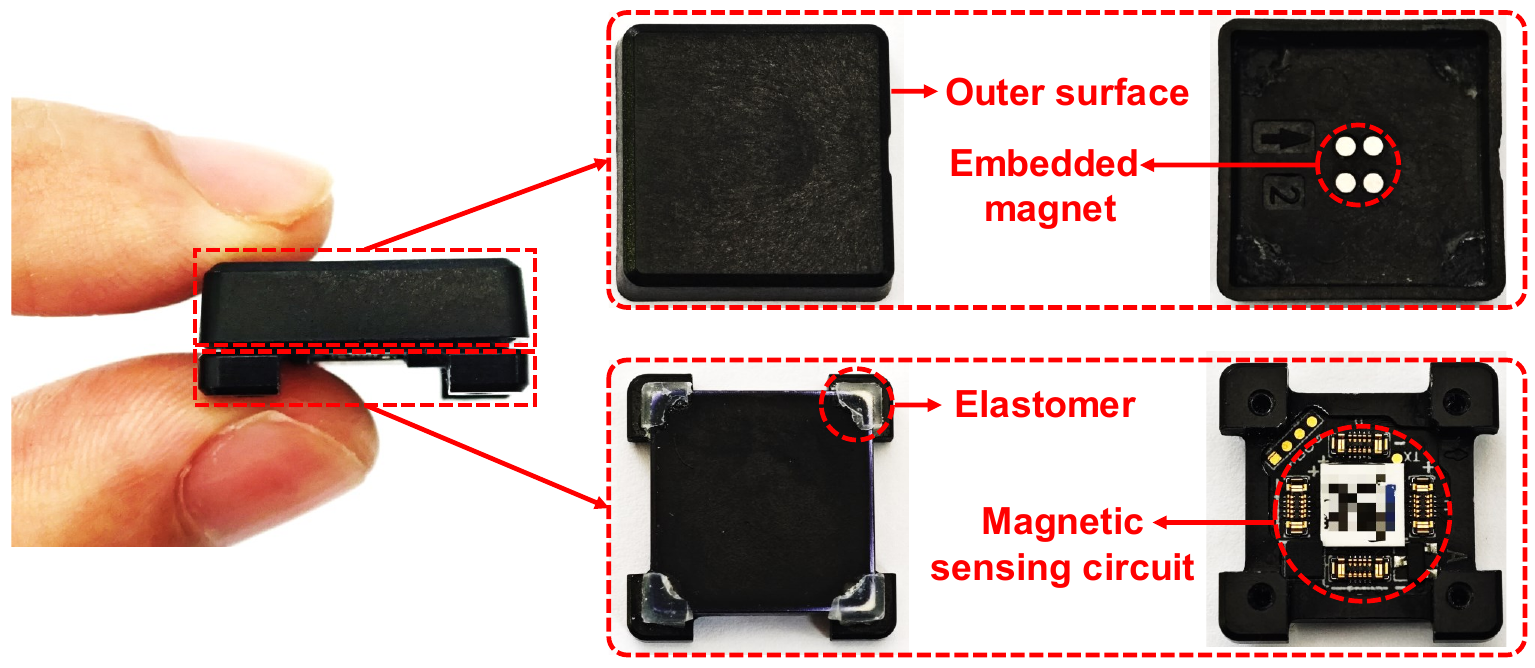}
    \Description{
    Internal view of a commercial magnetic tactile sensor,
    showing the sensor structure and internal components.
    }
    \caption{Inside view of PaXini PX6AX-GEN3-MC-M2020-Elite.}
  \end{subfigure}
  \caption{Principle and structure of magnetic tactile sensors. (a) demonstrates the neutral, compressed, and shear states. (b) Structure of a COTS magnetic tactile sensor.}
  \label{fig:background2}
\end{figure}

As embodied AI systems are deployed in increasingly complex physical environments, robotic manipulation has become critically dependent on multimodal sensing~\cite{zhao2025mpddm}. 
Among predominant perceptual modalities, e.g., vision and audition, tactile perception is fast emerging as an indispensable sensing ability for contact-rich manipulation~\cite{meribout2024tactile,bhirangi2025anyskin}. 
Tactile perception provides fine-grained measurements of contact states and force distributions that neither vision nor proprioception can fully replicate~\cite{andrussow2023minsight,mao2024multimodal}. 
Specifically, visual perception is susceptible to environmental interference such as lighting variations and occlusion, making it difficult to perceive contact states within visual blind spots; proprioception mainly reflects the robot's internal motion state and therefore cannot acquire fine-grained interaction data at the contact interface~\cite{luo2024robotics}.

In contrast, tactile perception demonstrates unique advantages in sensing granularity, real-time performance, and environmental resilience.
Tactile sensors can directly acquire the precise force feedback at the physical contact points. 
This core feature allows robots to accurately characterize key parameters~\cite{li2020tactile}, such as weight~\cite{silva2019mass}, stiffness~\cite{yuan2017hardness}, and friction coefficient~\cite{khamis2021friction}.
In the system scope, tactile sensing enables closed-loop force control systems, thus adjusting the robot's fine-grained manipulation in real time. 
It also prevents manipulation failures caused by perceptual deficits, thereby ensuring the safety and precision of human-robot collaboration.
This capability can be the cornerstone for several crucial real-world tasks.
For example, when a robotic gripper closes on an object, understanding subtle variations in the \emph{normal force} helps apply an appropriate force, thereby enabling a firm grip without crushing the object.
When a robotic hand is holding a fragile object while moving, sensing the \emph{shear force} in real-time can prevent the gripper from dropping or damaging the object.

To achieve tactile sensing, there are the following key approaches: optical, capacitive, resistive, and magnetic.
Specifically, vision-based tactile sensors are commonly composed of light sensors and an elastomer~\cite{gelsight2026official}. 
An applied force can incur deformation in the elastomer. 
Then, the light sensor can interpret the deformation, thus characterizing the force.
However, this approach is undermined by the cost of frame rate, image processing overhead, and packaging volume. 
For example, the GelSight Mini is listed at USD 510 and integrates an 8MP camera with a nominal frame rate of 25 FPS, while its package size is approximately 32~mm by 31~mm by 28~mm~\cite{gelsightMiniProduct,gelsightMiniDatasheet}.
Capacitive tactile sensors~\cite{singletact2026official} perform measurements via capacitance changes caused by force to decode tactile information.
However, this approach is susceptible to parasitic capacitance, noise, humidity, and temperature variations. 
Piezoresistive tactile sensors~\cite{tekscan2026official} utilize resistance changes caused by force to decode tactile information.
Although being a low-cost option, this line of approaches is prone to hysteresis, drift, and insufficient long-term stability. 

As we will elaborate in Sec.~\ref {sec:background}, magnetic tactile sensors are fast emerging as an approach that can achieve both multi-dimensional force sensing and low cost.
For example, magnetic tactile sensors can be composed of an underlying array of magnetometers and a magnetic elastomer.
As the magnetic field drops with distance cubed, subtle deformation of the elastomer can be captured by the sensor.
Compared to other tactile solutions, magnetic tactile sensing possesses significant advantages: rich three-dimensional force sensing, compact form factor, low cost, and resilience against light conditions. 

In this paper, we discover a fundamental physical vulnerability in magnetic tactile sensors.
That is, one can alter the environment's magnetic field, thus injecting erroneous force into magnetic tactile sensors.
As shown in Sec.~\ref {sec:background_limitation} and Sec.~\ref {sec:across_models}, this underlying vulnerability is present across multiple evaluated magnetic tactile sensor models.
Exploiting this vulnerability, we propose {\name} (as shown in Fig.~\ref{fig:overview of PhantomTac}), the first stealthy physical adversarial attack targeting magnetic tactile sensors using passive magnetic tags.

To achieve a fine-grained, stealthy attack on magnetic tactile sensors, it is crucial to answer the following key challenges:
(1) How to characterize commercial-off-the-shelf (COTS) magnetic tactile sensors? 
(2) How can the attack effectively alter/modulate the environment's magnetic field while staying stealthy? 
(3) How to affect the closed-loop force control, thus achieving severe damage to real-world objects?

Addressing the first challenge is essential for achieving an in-depth understanding of the magnetic tactile sensor's mechanism.
Specifically, magnetometers interpret the magnetic field as a three-dimensional field strength, i.e., $B_x, B_y$, and $B_z$.

Therefore, we conducted extensive reverse engineering research on representative COTS magnetic tactile sensors, e.g., {PX6AX-GEN3-MC-M2020-Elite}~\cite{paxini2026px6axgen3}. 
Specifically, we collected sensor response data from magnetic tactile sensors under the intervention of external magnetic sources, utilized this dataset to construct a fitting model based on the Radial Basis Function (RBF), and used this model to characterize the response characteristics of magnetic tactile sensors to external magnetic sources.

To address the second challenge, we employ micro-passive permanent magnets (with a thickness of 0.5 mm) as the attack source. 
This endows {\name} with extremely high stealth characteristics: it requires no external power supply, generates no monitorable thermal signals or acoustic signatures, and causes almost no changes to the visual appearance of the target object. Furthermore, since the magnetic field itself can penetrate non-ferromagnetic materials without attenuation, this attack is completely independent of line-of-sight conditions, allowing the attack tag to be well hidden under normal labels on the object's surface. Based on this, we can achieve the injection of forged normal or shear force signals through carefully designed attack tags while staying stealthy.

To address the third challenge, we characterize the target magnetic tactile sensor model's responses to external magnetic sources. 
This one-time-effort allows the attacker to customize the attack tag using the aforementioned passive magnetic tag design framework for deterministic force injection. 
Once deployed, when a humanoid robot or robotic arm performs a touching or grasping task, the attack tag prompts the sensor to generate erroneous raw force signals and feed them back into the closed-loop control system. 
{\name} also shows high robustness in repetitive experiments. For example, {\name} maintains an attack success rate above 95\% for grasping-position offsets within 8 mm in our experiments.
This signal injection causes the system to make erroneous judgments at the decision-making level, such as erroneously increasing or decreasing the contact force, ultimately leading to object damage or operational failure in the real world.
We demonstrated these end-to-end attacks in the online artifact.

In {\name}, we make the following contributions:
\begin{itemize}
    \item We discover and formally characterize an underlying hardware vulnerability in magnetic tactile sensors, namely the physical inseparability of internal and external magnetic flux, proving that current anti-interference mechanisms possess fundamental flaws when facing static adversarial magnetic fields.
    
    \item We propose {\name}, a passive and highly stealthy tactile perception attack framework. By utilizing reverse engineering, we design the passive magnetic tag that is as thin as a piece of paper.
    
    \item Based on the limitations of existing anti-magnetic interference mechanisms, we propose a novel defense mechanism to effectively mitigate such physical-layer security threats.
\end{itemize}

\section{Background}
\label{sec:background}

\subsection{Primer on Magnetic Tactile Sensing}
The rapid advancement of embodied intelligence, robotic manipulation, and dexterous grasping has underscored the critical importance of tactile perception. 
Compared with visual sensing, tactile feedback directly captures fine-grained contact states, force variations, and slip --- information essential for fine manipulation, closed-loop control, and operation in harsh environments (e.g., visual blind spots). 
Among emerging sensing modalities, magnetic tactile sensors have gained significant traction due to their high sensitivity, multidimensional force perception, and flexibility in integration.
In academia, the number of related research efforts has exhibited a rapid growth trend in recent years~\cite{man2024bionic,bhirangi2025anyskin,pattabiraman2025eflesh,hu2024magneticskin,yan2021softmagnetic}.
The industrial sector has also been heavily advancing magnetic tactile sensors since 2021.
Specifically, PaXini introduced various models (e.g., fingertip, palm, and finger pad) of magnetic tactile sensors in 2021~\cite{paxini2026official}.
Another leading manufacturer, XELA~\cite{xela2026official}, presented uSkin in 2021. 
Meta has open-sourced ReSkin~\cite{bhirangi2021reskin} in 2021.
Major semiconductor manufacturer, e.g., Melexis, introduced magnetic tactile sensors (Tactaxis) for a dexterous robotics system in 2022~\cite{le2022gradiometric}. 
In the following, we first describe the components of a magnetic tactile sensor. Then, we elaborate on its advantages. 

The fundamental working principle of magnetic tactile sensors is to convert mechanical forces 
into measurable magnetic field variations. 
As illustrated in Fig.~\ref{fig:background2} (a), the sensor typically consists of a flexible elastomer layer, an embedded magnet, and an underlying sensor array composed of several magnetometers, e.g., Hall-effect sensors~\cite{tomo2016hall} or magnetoresistive sensors~\cite{oh2017magnetic}. 
When an external pressure is applied to the elastomer, the elastic structure undergoes compressive, bending, or shear deformation, which alters the relative position, distance, or pose between the magnet and the magnetic sensing units.
As the magnetic field follows an inverse-cube law with the distance:
\begin{equation}
    \mathbf{B(r)}\propto\frac{1}{\mathbf{|r|}^3}
    \label{eq:proportion}
\end{equation}
Here, $\mathbf{B(r)}$ is the three-dimensional magnetic field strength at position $\mathbf{r}$, whereas $\mathbf{r}$ is the position of the magnet relative to the sensor,
$|\mathbf{r}|$ is the radial distance.
Therefore, the positional changes lead to variations in the local magnetic field distribution. 
Then, an analytical pipeline maps these magnetic field readings into tactile information such as contact location, normal force, shear force, and slip. 
For example, XELA and PaXini provided detailed charts~\cite{xela2026technology,paxini2026itpu} for converting the magnetic field to fine-grained multi-dimensional force information.

To better demonstrate the principles and structure of the magnetic tactile sensor, we disassembled a magnetic tactile sensor (PaXini PX6AX-GEN3-MC-M2020-Elite) to reveal the inside view, i.e., Fig.~\ref{fig:background2} (b).

\begin{table}[t]
    \centering
    \caption{Comparison of tactile sensing modalities.}
    \label{tab:tactile_comparison}
    \scriptsize
    \setlength{\tabcolsep}{3.5pt}
    \renewcommand{\arraystretch}{1.12}
    \resizebox{\columnwidth}{!}{%
    \begin{tabular}{lccccc}
        \toprule
        \textbf{Metric} & \textbf{Optical} & \textbf{Capacitive} & \textbf{Resistive} & \textbf{Magnetic} \\
        \midrule
        Fine-grained force sensing                       & \greencheck    & \redcross      & \redcross     & \greencheck \\
        Easy multi-dimensional force decoupling        & \greencheck    & \greencheck    & \redcross     & \greencheck \\
        Low cost \& overheads                            & \redcross      & \redcross      & \greencheck   & \greencheck \\
        Robust static \& dynamic sensing                 & \greencheck    & \greencheck    &\redcross      & \greencheck \\
        \bottomrule
    \end{tabular}%
        }
    \vspace{1pt}
    \parbox{\columnwidth}{\scriptsize * The symbols \greencheck{} and \redcross{} are used to indicate the relative ease of implementation or the typical sensing performance, rather than absolute feasibility.}
\end{table}

\subsection{Advantages of Magnetic Tactile Sensors}

\begin{figure*}[htbp] 
  \centering
  \includegraphics[width=0.95\textwidth]{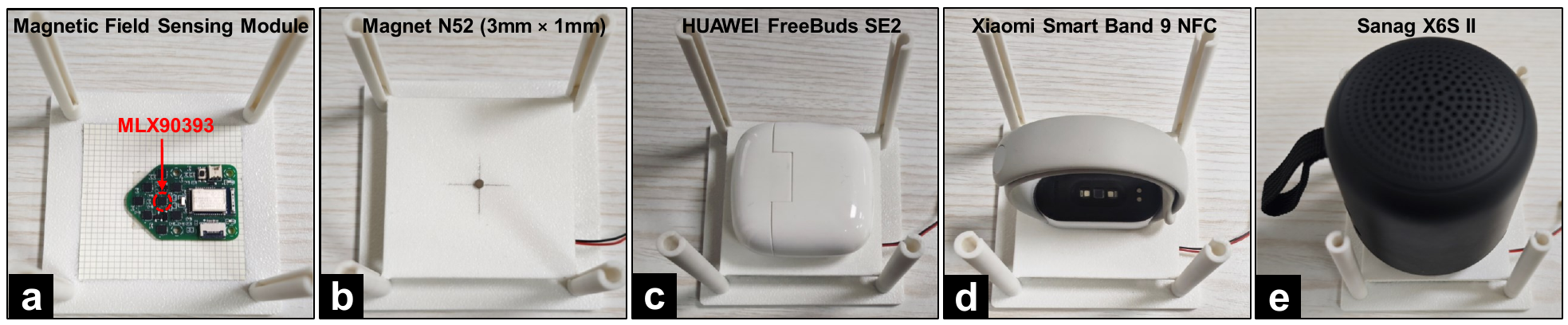} 
  \Description{Overview of existing robotic tactile sensing technologies and their underlying sensing principles.}
  \caption{Experiment on Anti-Interference Mechanisms of Magnetic Tactile Sensors. The platform for the sample is 10 mm away from the bottom surface of the sensor. (a) presents the placement of the MLX90393 Module; whereas (b)---(e) denote the placement of various magnetic objects.}
  \label{fig:background3}
\end{figure*}

\begin{figure}[t] 
  \centering
  \includegraphics[width=0.85\linewidth]{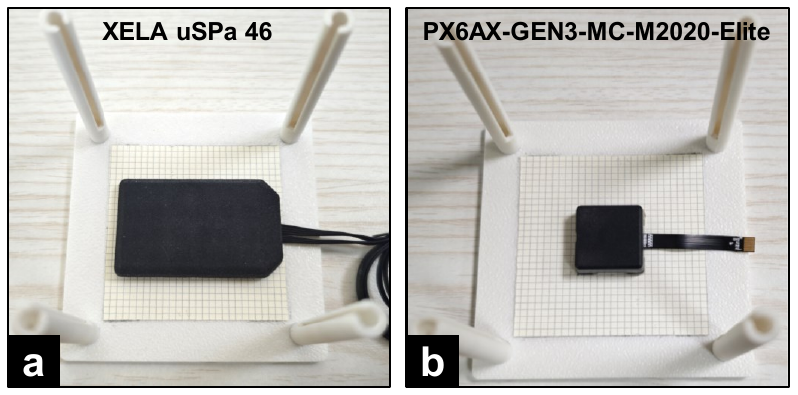}
  \Description{Example of a robotic tactile sensing system showing the sensor structure and the physical interaction process during contact.}
  \caption{Placement of PaXini and XELA Sensors. (a) XELA uSPa 46. (b) PX6AX-GEN3-MC-M2020-Elite.}
  \label{fig:background1}
\end{figure}

\noindent\textbf{Fine-grained force sensing.} 
Magnetic tactile sensors typically achieve perception by embedding magnets or magnetic materials within a flexible elastomer, e.g., silicone. 
When external contact induces deformation in the elastomer, the position and/or orientation of the magnet changes accordingly, thereby altering the local magnetic field distribution. Magnetic field sensors exhibit high sensitivity to such variations.
Consequently, even miniaturized displacements or slight contact forces can be effectively converted into measurable electrical signals, thereby enabling highly granular micro-force detection. 
For example, a single sensing point of the XELA uSkin sensor achieves a resolution of 0.1~gf, approximately equivalent to 0.98~mN~\cite{xela2026technology}.

\noindent\textbf{Multi-dimensional force decoupling.} 
The magnetic field can be characterized as a $1\times3$ vector in three dimensions, i.e., x, y, and z.
When forces are applied in different directions, they induce displacement or rotation of the magnet in distinct directions, thus leading to corresponding variations in the magnetic field. 
For example, normal forces predominantly manifest as magnetic field variations in the vertical direction to the magnetometer array. 
In contrast, shear forces are more readily reflected as magnetic field deviations in the horizontal plane of the magnetometer array. 
For example, the IPTU GEN3 series from PaXini~\cite{paxini2026px6axgen3} can interpret three-dimensional forces.

\noindent\textbf{Low cost \& overheads.}
The magnetometer array of magnetic tactile sensors is commonly comprised of 8-16 magnetometers --- tiny, chip-level sensors with an edge length of around 2 mm.
For each frame of the magnetic tactile data stream, the data is a $3\times Num$ matrix of single-precision (32-bit) floating-point numbers, thus minimizing the overhead in the peripheral readout.
These features lead to a compact form factor and low prices of magnetic tactile sensors.
For example, the price of the GelSight Mini (optical solution) is \$510~\cite{gelsightMiniProduct}, whereas the base kit of the SingleTact (capacitive solution) is around \$81~\cite{singletact2026calibration}.
In contrast, the PaXini PX-6AX GEN3 (magnetic solution) is around \$29 per unit~\cite{paxini2026px6axgen3mall}.
Thanks to the compact form factor, magnetic tactile sensors can also be integrated for various scenarios. 
For example, both PaXini~\cite{paxini2026official} and XELA~\cite{xela2026official} offer magnetic tactile sensor solutions tailored for various locations, e.g., fingertips, finger pads, and palms.

\noindent\textbf{Robust static \& dynamic sensing.} 
Regarding static perception, magnetic tactile sensors are typically capable of directly measuring absolute magnetic fields or differential magnetic field signals. Therefore, they are more adept at stably characterizing the force state under continuous contact conditions. Regarding dynamic perception, their advantages primarily stem from the high sampling frequency of the underlying magnetic field chips, as well as the high sensitivity of the magnetic field to the displacement and rotation of the magnet. Taking the public data of the PaXini PX-6AX GEN3 as an example, its sampling frequency reaches up to 1,000,000~Hz, with a maximum output frequency of 1000~Hz~\cite{paxini2026itpu}. The 2023 paper by the Melexis team~\cite{LeSignor2023MassManufacturable} also indicates that magnetic tactile sensors possess high-speed three-dimensional perception capabilities, demonstrating their potential for slip detection applications in industrial grippers.

In Table ~\ref{tab:tactile_comparison}, we characterized the key features of optical, capacitive, resistive, and magnetic tactile sensors. 
However, to build practical systems based on these key features, it is also essential to mitigate the impact of common interferences, e.g., the magnetic field incurred by the Earth and everyday electronics.
In the following, we will elaborate on existing efforts to overcome interference and highlight their limitations. 

\subsection{Existing Methods for Mitigating Magnetic Interferences}
\label{sec:background2.3}
In a magnetic tactile sensor, the magnetic field sensed by a magnetometer, i.e., $\mathbf{B}$ can be denoted as the following formula:
\begin{equation}
\mathbf{B} = \mathbf{B_{em}} + \mathbf{B_{E}} + \mathbf{B_{others}}
\label{eq:denoise}
\end{equation}
Here, $\mathbf{B_{em}}$ denotes the magnetic field of the embedded magnet, $\mathbf{B_{E}}$ and $\mathbf{B_{others}}$ denote the Earth's and other interfering magnetic fields (e.g., incurred by electronic devices), respectively.
Hence, to improve the reliability of a magnetic tactile sensor, the core is removing the interfering components, i.e., $\mathbf{B_{E}} + \mathbf{B_{others}}$.

For a single magnetometer, it is infeasible to differentiate $\mathbf{B_{em}}$ from other components.
To overcome this challenge, Le Signor et al.~\cite{le2022gradiometric} proposed a gradiometric design that uses multiple closely aligned magnetometers within a single compact IC package (about 2~mm span).
Now, based on Eq.~\ref{eq:denoise}, we can denote sensor $i$'s measurement of the magnetic field as:
\begin{equation}
\mathbf{B_i} = \mathbf{B_{em, i}} + \mathbf{B_{E}} + \mathbf{B_{others, i}}
\end{equation}
Note that the magnetic field strength shows cubic attenuation over distance as shown in Eq.~\ref{eq:proportion}.
Hence, the magnetic field distribution incurred by electronics (e.g., earphones) demonstrates an approximately uniform distribution among these sensors.
In contrast, the embedded magnet's distance is only a few millimeters away.
Hence, by subtracting the readings of adjacent magnetometers, one can effectively remove the approximately uniform components of $\mathbf{B_{E}} + \mathbf{B_{others}}$, while preserving the spatial variation of $\mathbf{B_{em}}$. 
This suppression relies on the assumption that the external interfering field shows an approximately uniform distribution across magnetometers.
This approach has been adapted by major tactile sensor manufacturers such as Melexis~\cite{le2022gradiometric}.

\subsubsection{Limitations}
\label{sec:background_limitation}
\begin{figure}[t] 
  \centering
  \includegraphics[width=0.97\linewidth]{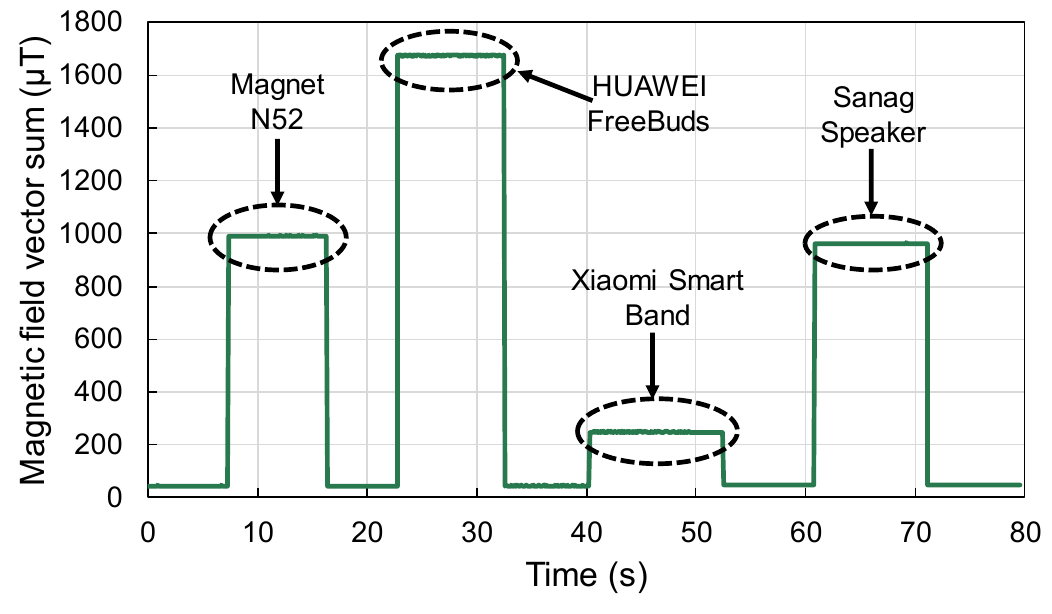}
  \Description{Illustration of magnetic tactile sensing mechanisms showing the relationship between magnetic fields and tactile sensor responses.}
  \caption{Comparison of magnetic field intensity between common magnetic objects in daily life and small magnets.}
  \label{fig:background4}
\end{figure}

\begin{figure}[t] 
  \centering
  \includegraphics[width=1.02\linewidth]{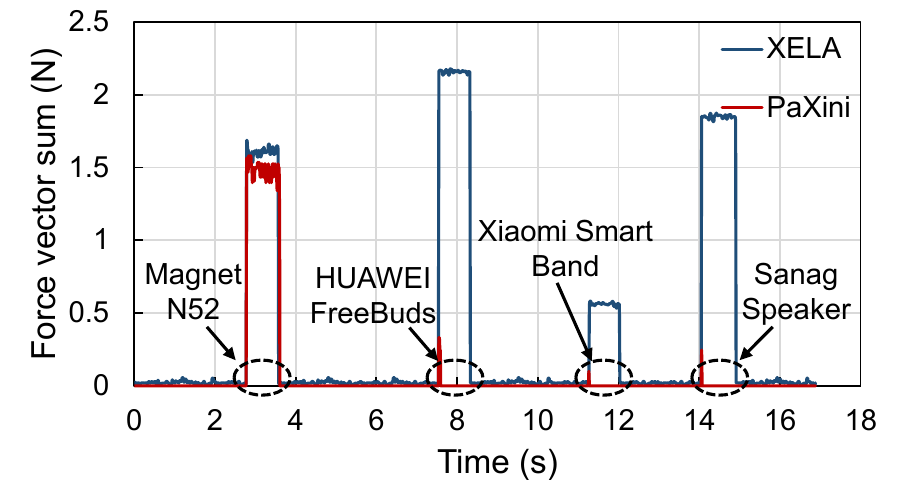}
  \Description{Comparison of different tactile sensing modalities highlighting the characteristics and advantages of magnetic tactile sensors.}
  \caption{Comparison of interference on magnetic tactile sensors caused by different objects.}
  \label{fig:background5}
\end{figure}

Although the aforementioned approach can effectively remove $\mathbf{B_{E}}+\mathbf{B_{others}}$, there is a major loophole:
If the interfering magnetic field is not uniformly distributed across the magnetometers, the subtraction approach would fail.
Then we ask a question: Is it possible to produce such a magnet in the real world?
The most intuitive interfering source is a small, strong magnet placed close to the underlying sensor array. Due to the short distance, its magnetic field varies across adjacent magnetometers instead of appearing as a uniformly distributed interference. 
Therefore, subtraction cannot fully cancel this external magnetic field. 
The remaining differential signal may therefore be interpreted as a change in the embedded magnet's field and converted by the tactile sensor into an erroneous force reading.

To verify this, we selected up-to-date COTS magnetic tactile sensor models, i.e., PaXini PX6AX-GEN3-MC-M2020-Elite and XELA's uSPa 46 (as shown in Fig.~\ref{fig:background1}), to perform a real-world study.
For the interfering magnetic field, we select representative electronic devices (e.g., earbuds and speakers) and a small passive magnet.
Here, we use electronic devices to verify if the tactile sensors are competent for overcoming common interferences.
We picked the passive magnet for verifying the aforementioned loophole.

We designed a comparative experiment (as shown in Fig.~\ref{fig:background3}) that evaluated sensor behavior under two categories of external magnetic input. 
Specifically, we categorized the test sources into two classes: ambient magnetic field sources and localized concentrated magnetic objects. 
The ambient stray magnetic field sources encompass the natural background magnetic field and several everyday electronic devices (e.g., smart bands, Bluetooth earphones, and Bluetooth speakers); whereas localized concentrated magnetic objects (e.g., small strong magnets) are utilized to simulate external magnetic disturbances with higher spatial locality and more pronounced magnetic field gradients. 

In this experiment, we maintained the installation configuration and relative testing posture of the magnetic tactile sensors consistent, exclusively varying the category of the external magnetic source (as shown in Fig.~\ref{fig:background3}(b)---(e)). 
We first employed the MLX90393~\cite{melexis_mlx90393_datasheet} to record the magnetic field of each object (as shown in Fig.~\ref{fig:background3}(a)).
This step is for verifying whether the object's magnetic field strength is sufficient.
Next, we recorded the sensor outputs by putting the object close to the magnetic tactile sensor (as shown in Fig.~\ref{fig:background3}(b)---(e)).
This step is for verifying whether the object's magnetic field can interfere with the tactile sensor.
The distance is fixed to be 10 mm.

As shown in Fig.~\ref{fig:background4}, all objects demonstrate strong magnetic field strength. Note that the earbuds' magnetic field is even stronger than the passive magnet.
However, Fig.~\ref{fig:background5} demonstrated that for the PaXini sensor, electronic devices do not have a sustained effect on the tactile sensor. 
Here, the force values in the figure represent the vector sum of $F_x$, $F_y$, and $F_z$.
This experiment revealed two key findings: 
(1) The small passive magnet produces erroneous readings on both major tactile sensors.
(2) The PaXini model achieves better resilience against an interfering magnetic field compared with XELA's model. (That is, under magnetic interference from common electronic devices, the abnormal output of XELA is 5.8--7.7 times that of PaXini, and the abnormal output of XELA persists continuously.)

This experiment demonstrates that a small passive magnet without sensor-specific customization can effectively interfere with multiple COTS magnetic tactile sensor models, indicating that the underlying vulnerability is not limited to a single sensor model.
Next, we ask: Is it possible to formulate a practical attack for these tactile systems?

\section{Attack Model and Challenges}

\subsection{Attack Model}
The attacker's goal is to stealthily interfere with the magnetic tactile sensors of a robotic system, e.g., a robotic arm or humanoid robot.  
By corrupting its force perception, the attacker can hijack the closed-loop force feedback control and trigger severe physical consequences, including dropping or crushing the targeted objects (as shown in Fig.~\ref{fig:overview of PhantomTac}). 
We make the following assumptions for the attack: 

\begin{figure}[t] 
  \centering
  \includegraphics[width=\linewidth]{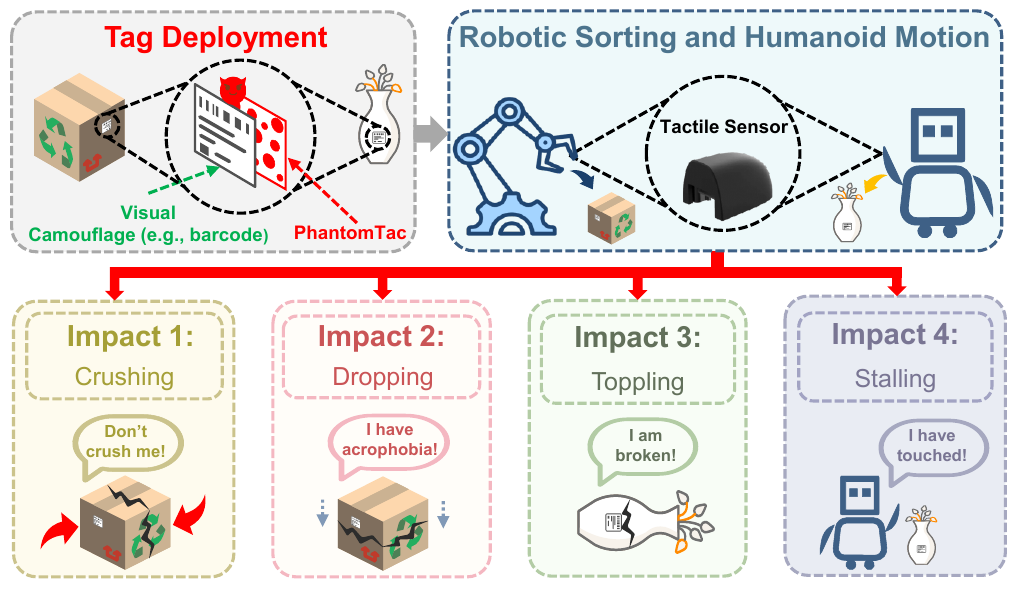}
  \Description{Overview of the PhantomTac attack framework showing the attacker, deploy-and-forget magnetic tags, robotic tactile sensors, and the resulting manipulation of robotic perception.}
  \caption{An overview of {\name}'s attack model.}
  \label{fig:overview of PhantomTac}
\end{figure}

\noindent\textbf{No system intrusion required.} 
{\emergencystretch=1em
We assume that the attacker only needs brief physical access to a target object or its packaging to deploy a magnetic tag near an expected contact region. Such access can occur in scenarios such as logistics sorting, malicious parcel delivery, or shared human--robot environments. 
No access to the robot's control system or software is required.
\par}

\noindent\textbf{Deploy-and-forget.} 
We assume that the attacker requires only a single, one-time deployment to implement the attack. 
The tag can be deployed in advance, and the attacker does not need to stay close to the target.
The attack tag is a compact array of small permanent magnets and can be camouflaged as QR codes, trademarks, or even stickers. 
As we will present in Sec.~\ref{sec:design}, the tag can be as small as 1.44 $cm^2$ and uses miniature magnets with a thickness of 0.5 mm.
Once the tactile sensor contacts the tag, the sensor will be interfered with by the tag, thus requiring no further attacker involvement.

\noindent\textbf{Knowledge of target sensor specifications.} 
We assume that the attacker knows the brand and model of the target sensor. Specifications of tactile sensors deployed on commercial robots are often publicly available. 
Such knowledge can be further refined by purchasing a sensor of the same model and performing offline characterization to quantify its magnetic vulnerabilities. This characterization requires no access to the victim robot, controller, software, runtime data, communication interfaces, or deployed tactile sensor.

\subsection{Key Challenges}
\noindent\textbf{Designing a stealthy and passive attack tag.} 
Existing physical-layer injection attacks typically rely on bulky active signal generators or electromagnetic coils. 
These require external power and are easily detected due to the bulky form factor.
In contrast, {\name} utilizes purely passive permanent magnets as attack tags, overcoming the constraint of power sources.
Now, the challenge lies in how to design tags that are compact, yet capable of generating sufficient magnetic flux to deceive high-sensitivity tactile sensors.
Note that to conceal a {\name} tag, the attacker may cover it with a camouflage label (e.g., a sticker).
Therefore, it is essential to assure this extra thickness would not impede the attack's efficacy.
Achieving these goals can make {\name} even stealthier while assuring the attack's performance.

\noindent\textbf{Achieving deterministic force injection.} 
As elaborated in Sec.~\ref{sec:background}, magnetic tactile sensors commonly integrate \emph{differential sensing arrays} to overcome magnetic interference. 
As we presented in Sec.~\ref{sec:background_limitation}, a small, passive magnet can disrupt the sensor reading.
However, it is challenging to achieve deterministic force injection. 
Meeting this objective is essential for fine-grained shear- and normal-force injection. Such injection can induce loose or overly tight grasps of objects such as a fragile vase.

\noindent\textbf{Handling contact-position variations.}
In practical robotic manipulation, the actual contact position may show subtle variations. 
{\name} therefore needs to remain effective under the contact-position variations. In structured manipulation tasks, object geometry and grasp planning can constrain the likely contact region, making tag placement more predictable. We evaluate this robustness in detail in Sec.~\ref{sec:Robustness}.

\section{Vulnerability Analyses}
\label{sec:vulnerability}
\subsection{Reverse Engineering of Magnetic Tactile Sensors}
\label{sec:reverse engineering}
To address the aforementioned questions, it is critical to understand the specifications of COTS magnetic tactile sensors. 
We propose an offline, model-specific \emph{reverse engineering approach}. This process can be performed on a separately obtained sensor of the same model and requires no access to the robotic system.
Specifically, we construct a three-dimensional testing space on the contact interface of the magnetic tactile sensor, and utilize small passive magnets to incur the interference.
This procedure reveals how the tactile sensor would generate force readings given various magnetic field settings.

\begin{figure}[t] 
  \centering
  \includegraphics[width=\linewidth]{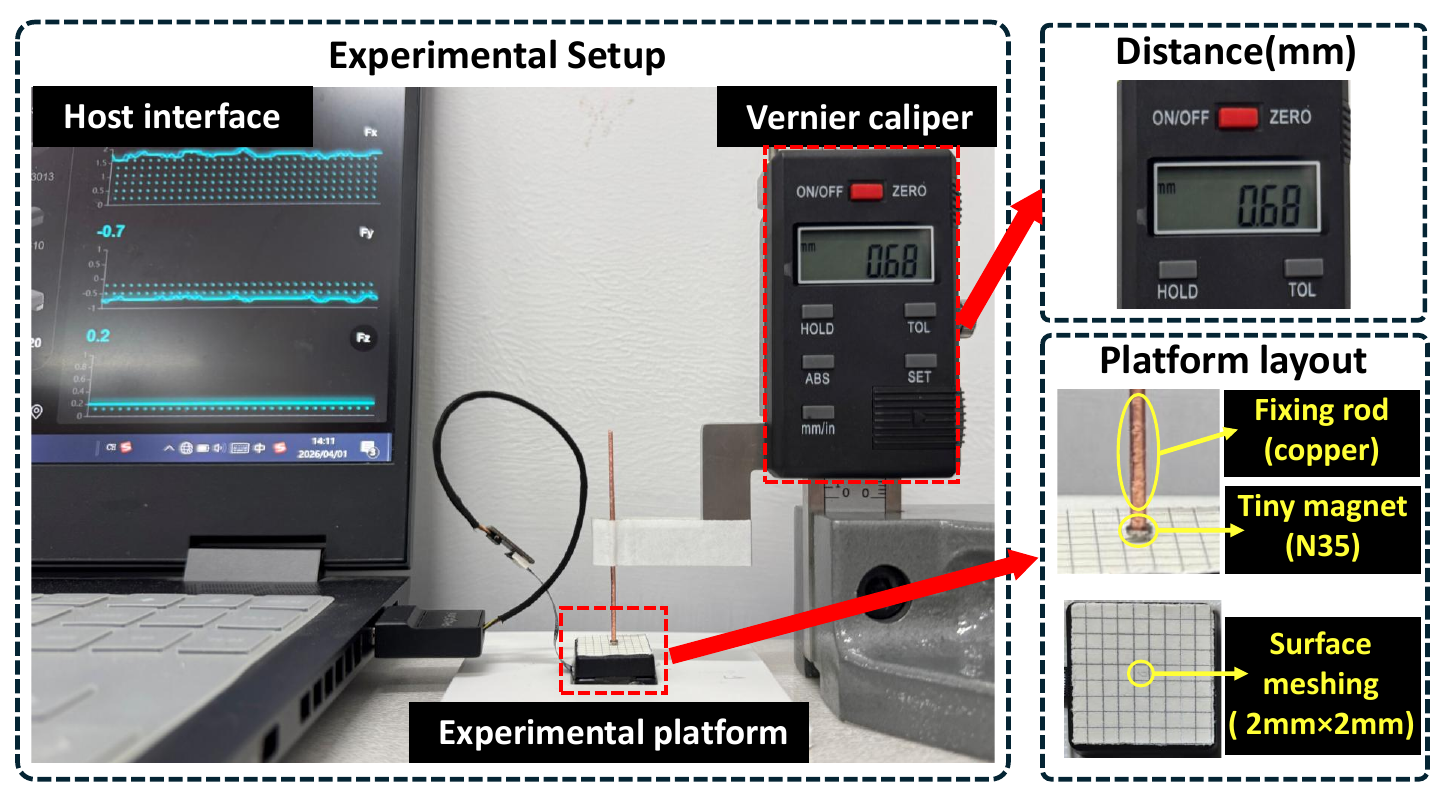}
  \Description{Logical diagram illustrating the reverse engineering process of PhantomTac, including the magnetic tactile sensor, measurement setup, and the procedure used to characterize sensor responses for constructing the attack.}
  \caption{Setup for the reverse engineering process.}
  \label{fig:vulnerability1}
\end{figure}

\begin{figure*}[htbp] 
  \centering
  \includegraphics[width=0.95\textwidth]{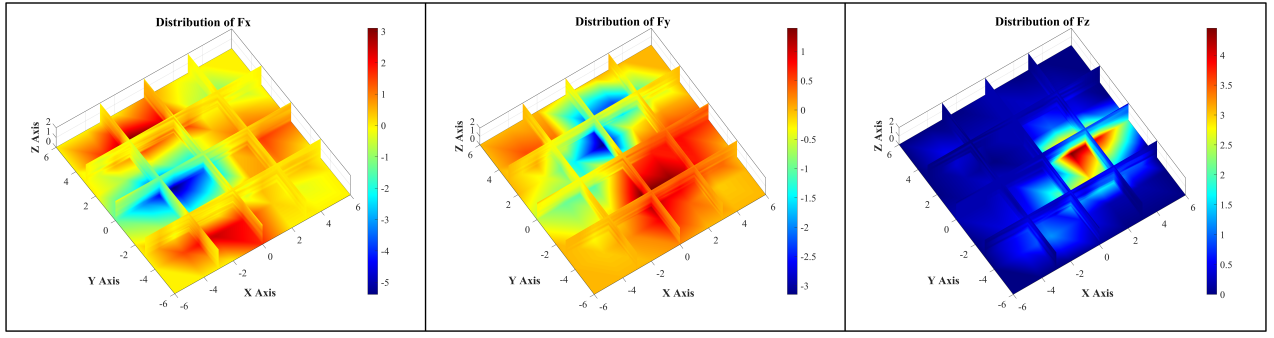} 
  \Description{Three-dimensional force feedback visualization of a magnetic tactile sensor, showing the sensor response under magnetic interference from a 3 mm diameter magnetic source.}
  \caption{Three-dimensional force feedback of magnetic tactile sensor. The diameter of the magnetic interference source is 3 mm.}
  \label{fig:vulnerability2}
\end{figure*}

As shown in Fig.~\ref{fig:vulnerability1}, we used the PaXini palm tactile sensor (PX6AX-GEN3-MC-M2020-Elite) for reverse engineering because of its resilience to magnetic field interference.
With the geometric center of the sensor's contact interface as the origin point, we defined a three-dimensional testing space.
The plane of this testing space was partitioned using grid paper with 2 mm × 2 mm squares, and its coverage area is 12 mm × 12 mm, i.e., ranging from -6 mm to 6 mm along the X and Y axes, with a planar step size of 2 mm.
The vertical space was partitioned using a vernier caliper (Nscing Es 889-101~\cite{ebuy7_nscing_889101}, precision 0.01 mm).
The vertical height ranges from 0 mm to 2.0 mm along the Z axis, with a vertical step size of 0.25 mm.
With this procedure, we set a 12 mm × 12 mm × 2 mm three-dimensional testing space above the sensor's contact interface.
These settings are depicted in Fig.~\ref{fig:vulnerability1}.
For the magnetic interferences, we used two specifications of cylindrical magnets.
Specifically, we used N35-grade passive magnets with a thickness of 0.5 mm and diameters of 2.5 mm and 3 mm.
Next, we moved the magnet to different positions atop the tactile sensor and recorded the force readings, i.e., ($F_x, F_y, F_z$).
This process yields a dataset of 882 discrete force-feedback measurements.
Based on this dataset, we constructed spatial heatmaps for the three-dimensional forces. 
As shown in Fig.~\ref{fig:vulnerability2}, these heatmaps illustrate the distribution of the tactile sensor's response to external magnetic stimuli. 

\subsection{Design of Attack Modalities}

After obtaining the response characteristics of the magnetic tactile sensor to external magnetic stimulation, we further conducted the attack design of the magnetic tag. 
To achieve an effective attack on the magnetic tactile sensor, the tag design needs to simultaneously satisfy the following conditions:

\noindent\textbf{(1) Balancing stealth and the magnetic field strength.} 
The tag must maintain a minimal form factor (size and thickness) to avoid visual detection or mechanical interference during deployment. 
However, the volume must remain sufficient to generate a magnetic flux density capable of inducing a measurable tactile response.

\noindent\textbf{(2) Decoupled force injection.} The design must allow for the independent manipulation of $F_x, F_y,$ and $F_z$.
The robotic control system interprets shear forces ($F_x$ and $F_y$) and normal forces ($F_z$) differently.
For example, a robotic gripper can use the shear force variations for detecting slip.
Therefore, the attacker may wish to inject significant tangential anomalies while suppressing $F_z$.

To achieve this, we identify high-sensitivity regions for specific force components. By strategically positioning a miniature magnet at a coordinate where the sensor exhibits a strong response to the target component and minimal response to others, we maximize the fidelity of the injected force signal.
While heatmaps provide a coarse distribution, we perform spatial interpolation modeling in Sec.~\ref{sec:design} to derive continuous response functions, enabling the precise localization of optimal injection points.

As we will show in Sec.~\ref{sec:design}, with this design principle, we can inject a normal force of approximately 4.2 N into the magnetic tactile sensor with only one thin passive magnet (with diameter of 3 mm). This magnitude of false force feedback is already sufficient to cause gripping force regulation anomalies in real-world robotic systems, such as excessively loose or excessively tight grasping. Meanwhile, the impact of this magnet on the force components in the other two directions does not exceed 0.7 N.

\subsection{Improving the Attack Stealthiness}
To achieve the \emph{deploy-and-forget} capability, it is essential to ensure stealthiness. 
In {\name}, we selected miniature magnets with a thickness of 0.5 mm as the attack source. 
Miniature magnets of this size can be concealed beneath camouflage labels without exhibiting obvious visual anomalies.

\begin{table}[t]
    \centering
    \caption{Common Sticker Thicknesses.}
    \label{tab:label_thickness}
    \resizebox{\columnwidth}{!}{%
    \begin{tabular}{llc}
        \toprule
        \textbf{Label Type} & \textbf{Common Applications} & \textbf{Thickness (mm)} \\ 
        \midrule
        Thermal Paper~\cite{avery_direct_thermal_200gp} & Express shipping labels & $0.142 \pm 10\%$ \\
        Coated Paper~\cite{avery_high_gloss_paper} & Product barcodes & $0.183 \pm 10\%$ \\
        Synthetic Paper~\cite{avery_synthetic_paper_bw0153} & Waterproof QR code labels & $0.145 \pm 10\%$ \\
        PET Label~\cite{avery_pet_sy7201} & Electronic product back labels & $0.115 \pm 10\%$ \\
        \bottomrule
    \end{tabular}%
    }
\end{table}

As shown in Table~\ref{tab:label_thickness}, common logistics and product labels typically range from $0.1$ to $0.2$~mm in thickness. 
Because the non-ferromagnetic substrates (e.g., paper, adhesive, PET) have a relative magnetic permeability near unity ($\mu_r \approx 1$), 
this thin camouflage does not impede the efficacy of the magnetic tag underneath.

We performed an experiment to verify this. 
Specifically, we performed a vertical scan using the 2.5 mm diameter magnet --- our smallest one. 
By constraining the planar coordinates $(x, y)$ and treating the $Z$-axis elevation as the variable, we profiled the attenuation profiles of the induced $F_x, F_y$, and $F_z$ components. 
Based on the spatial heatmaps (Fig.~\ref{fig:vulnerability2}), we selected $(0, 0, z)$, $(0, 2, z)$, and $(2, 0, z)$ as the data extraction trajectories to characterize the force-distance relationship.

\begin{figure}[t] 
  \centering
  \includegraphics[width=0.97\linewidth]{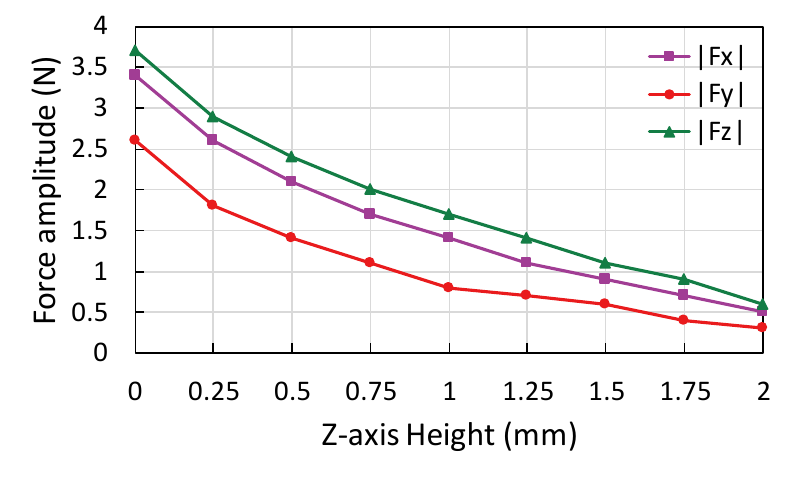}
  \Description{Experimental evaluation of the penetration capability of a 2.5 mm diameter magnet, showing the effect of magnetic interference through different distances or material layers on the tactile sensing response.}
  \caption{The penetration capability of 2.5 mm magnet.}
  \label{fig:vulnerability3}
\end{figure}

\begin{figure}[t] 
  \centering
  \includegraphics[width=1\linewidth]{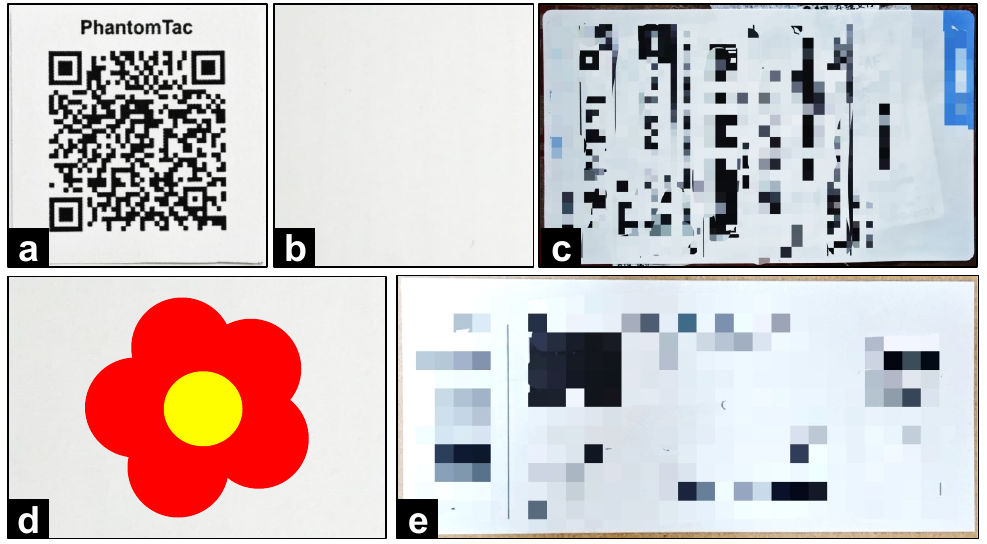}
  \Description{Examples of disguised appearances of PhantomTac tags. The subfigures show the attack device embedded or concealed as a QR code, a blank paper label, a shipping label, a cartoon sticker, and a product sticker.}
  \caption{Disguised appearance of {\name}. (a)---(e) present QR code, blank paper, shipping label, cartoon sticker, and product sticker, respectively.}
  \label{fig:vulnerability4}
\end{figure}

As shown in Fig.~\ref{fig:vulnerability3}, the magnet maintains a stable signal penetration at a vertical offset of $1$ mm,
thus significantly exceeding the thickness of standard labels. 
Note that higher-grade magnets (e.g., N52) would further extend this effective range. 
As demonstrated in Fig.~\ref{fig:vulnerability4}, the resulting {\name} tags are indistinguishable from ordinary stickers, allowing them to remain visually concealed before robotic contact. 
This stealthiness distinguishes {\name} from conventional sabotage (e.g., applying a slippery substance or directly damaging the object). 
Specifically, although conventional sabotage can lead to direct damage, this approach requires unrestricted physical access and can be easily noticed by the victim. 
In contrast, {\name} can instead remain concealed until the tagged object contacts the tactile sensor. 
Therefore, even if the tag is discovered after an incident, object damage or workflow disruption has already occurred.

\section{Magnetic Tag Design and Fabrication}
\label{sec:design}
\subsection{Inverse Magnetic Design}
\label{sec:design1}
To precisely localize the response-sensitive regions of the magnetic tactile sensor across $F_x, F_y,$ and $F_z$, we first performed interpolation of the aforementioned dataset.
Given that each discrete sampling point $(x, y, z, F_x, F_y, F_z)$ represents a physical state, we model the force reading as a unified vector output.
Based on these data-coupling requirements, we employed the Radial Basis Function (RBF) interpolation technique to construct a continuous force-field mapping model, $F = f(x, y, z)$. 
Given a set comprising $N$ spatially discrete sampling points $\{\mathbf{x}_i\}_{i=1}^N$, the three-dimensional force field predictive output at an arbitrary target spatial coordinate $\mathbf{x} = (x, y, z)$ can be defined as:

\begin{equation}
f(\mathbf{x}) = \sum_{i=1}^{N} \mathbf{w}_i \phi(\|\mathbf{x} - \mathbf{x}_i\|)    
\end{equation}

Here, $\|\mathbf{x} - \mathbf{x}_i\|$ denotes the Euclidean spatial distance, and $\phi(\cdot)$ represents the basis function.
In this paper, we adopt the Thin Plate Spline (TPS) kernel function, $\phi(r) = r^2 \ln(r)$. $\mathbf{w}_i \in \mathbb{R}^3$ is derived as three-dimensional vectors.
This approach enables us to jointly interpret the shear force and normal force at the same coordinate.

To evaluate the predictive robustness of the reconstructed response model, we introduced a Monte Carlo cross-validation procedure in the model construction phase.
In each of the 1000 iterations, 3\% of the dataset was randomly selected as the test set, while the remaining 97\% was used for the training process.

To comprehensively evaluate the predictive goodness-of-fit for the three-dimensional force in the blind tests, this study introduces three evaluation metrics: predictive performance score, median score, and 50-iteration moving average. 
When calculating the predictive performance score, we first clip the model outputs to their respective valid measurement ranges, i.e., the range for $F_x$ and $F_y$ is $[-10.0, 10.0]$, and the range for $F_z$ is $[0, 25.0]$. 
Subsequently, we compute the coefficient of determination (i.e., $R^2$) independently for the three coordinate axes, $F_x$, $F_y$, and $F_z$. 
The final score is defined as the arithmetic mean percentage of the non-negative $R^2$ values across the three axes, namely:

\begin{equation}
\text{Score} = \frac{1}{3} \sum_{i \in \{x,y,z\}} \max(0, R^2_i) \times 100\% 
\end{equation}

\begin{figure}[t]
  \centering
  \begin{subfigure}{0.48\linewidth}
    \centering
    \includegraphics[width=1.05\linewidth]{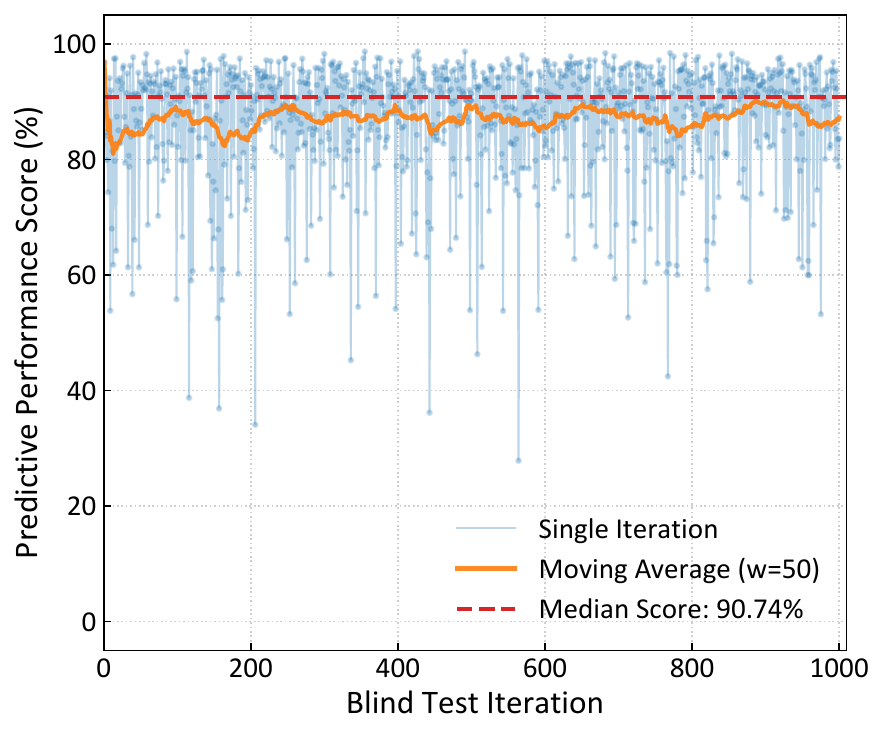}
    \Description{
    Blind test results for 3 mm diameter magnets, showing the measured responses or classification outcomes of the tactile sensing system under different test samples.
    }
    \caption{}
  \end{subfigure}
  \hspace{1pt}  %
  \begin{subfigure}{0.48\linewidth}
    \centering
    \includegraphics[width=1.05\linewidth]{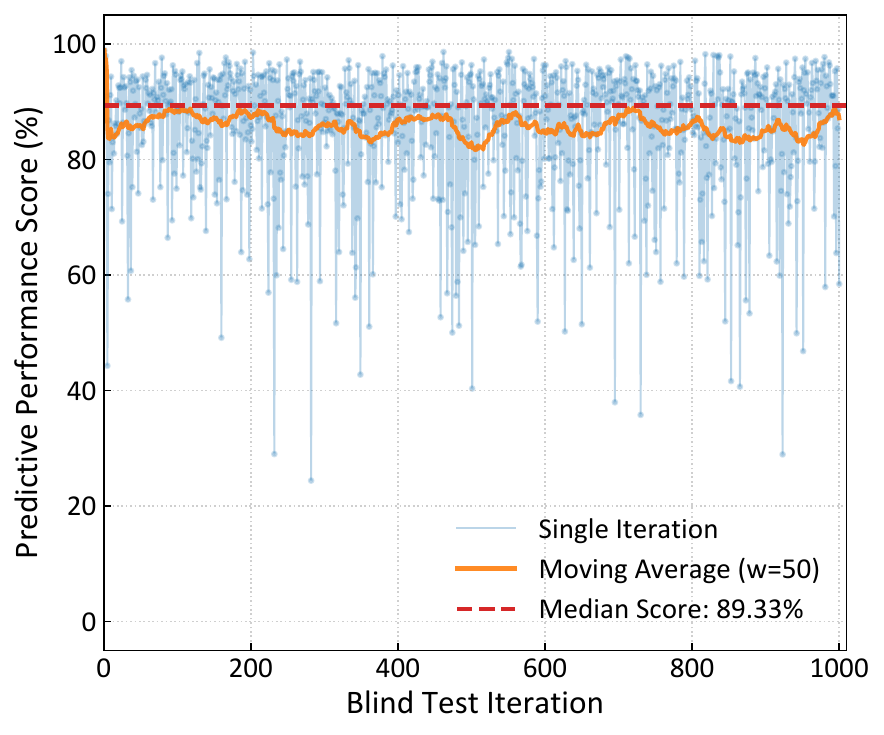}
    \Description{
    Blind test results for 2.5 mm diameter magnets, showing the measured responses or classification outcomes of the tactile sensing system under different test samples.
    }
    \caption{}
  \end{subfigure}
  \caption{Blind test results with magnets of different diameters. (a) Blind test results of 3 mm diameter magnets. (b) Blind test results of 2.5 mm diameter magnets.}
  \label{fig:design1}
\end{figure}

As shown in Fig.~\ref{fig:design1}, the model exhibited high robustness.
The median predictive scores for the 3~mm and 2.5~mm magnet datasets reached $90.74\%$ and $89.33\%$, respectively.  
The mean absolute errors (MAE) remained below 0.09~N, while the root mean square error (RMSE) was around 0.16~N. 
The 50-iteration moving average remained stable within the 80\% --- 90\% range.
These results indicate that the model achieves stable predictive performance without substantial overfitting on the sampled response data.

\begin{figure*}[htbp] 
  \centering
  \includegraphics[width=0.9\textwidth]{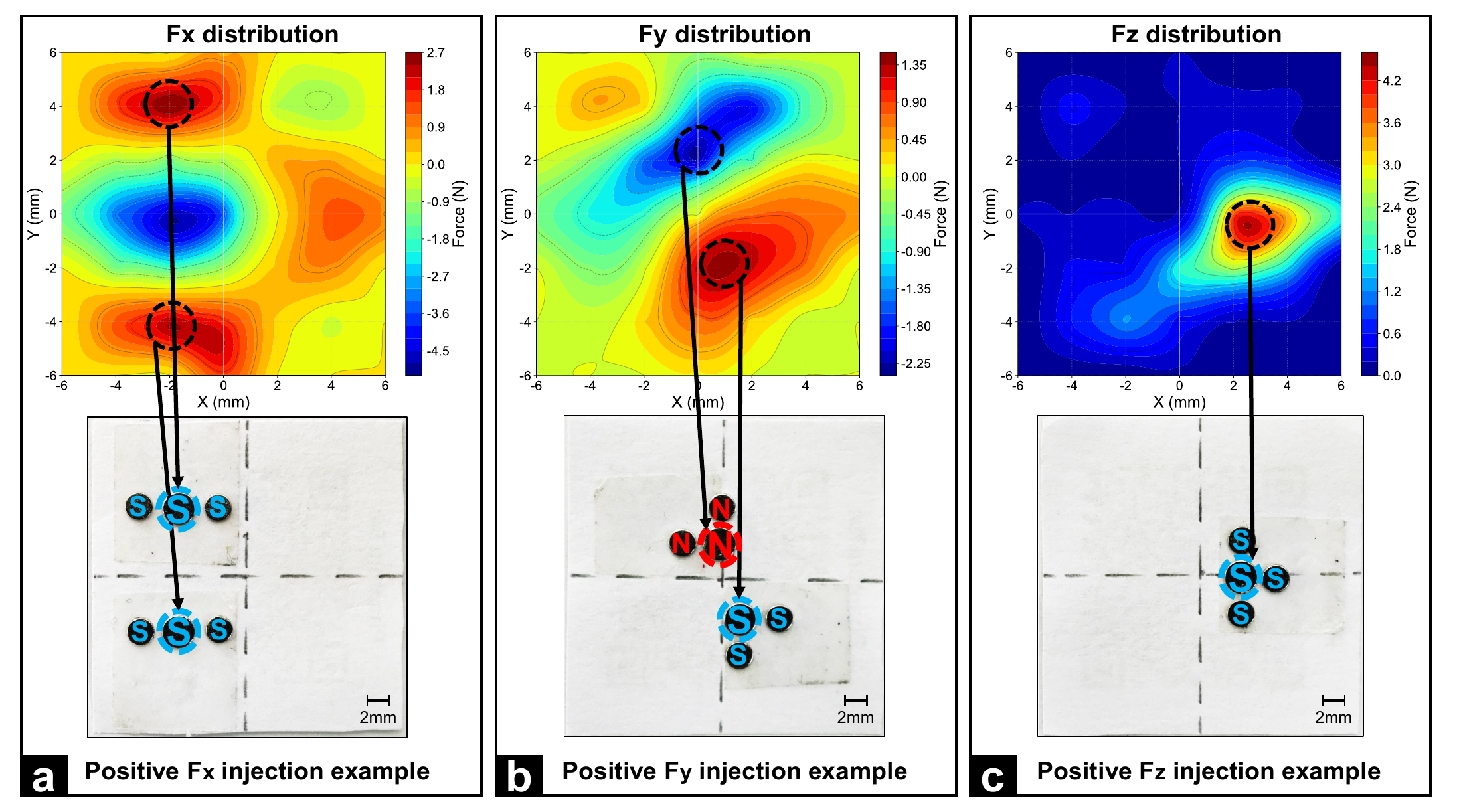} 
  \Description{Fabrication process of PhantomTac attack tags based on contour comparison of magnetic tactile sensor force-feedback patterns, showing how sensor response contours are analyzed and used to construct effective attack tags.}
  \caption{Fabricate attack tags based on the contour comparison of magnetic tactile sensor feedback force.}
  \label{fig:design2}
\end{figure*}

\begin{figure}[t]
  \centering
  \begin{subfigure}{0.485\linewidth}
    \centering
    \includegraphics[width=\linewidth]{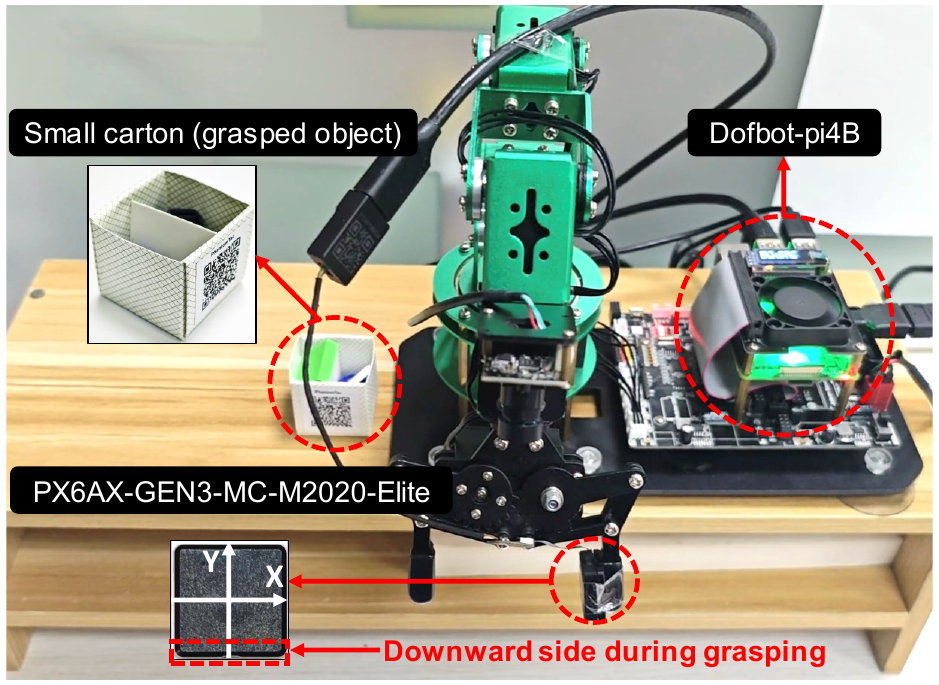}
    \Description{
Test setup for robotic grasping on the Dofbot-pi4B platform with a magnetic tactile sensor mounted on the inner side of the gripper.
}
    \caption{}
  \end{subfigure}
  \hspace{1pt}  %
  \begin{subfigure}{0.485\linewidth}
    \centering
    \includegraphics[width=\linewidth]{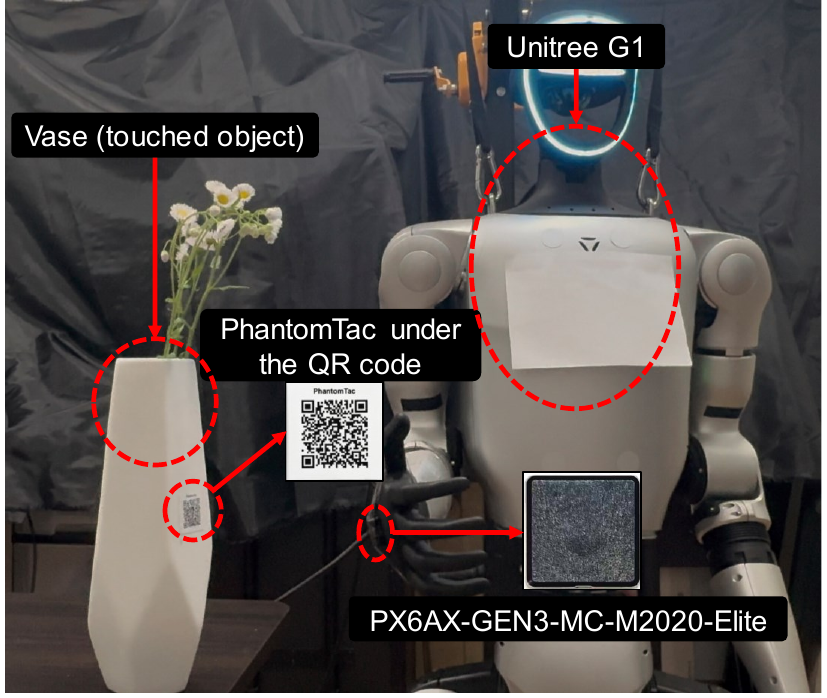}
    \Description{
Test setup for the Unitree G1 humanoid robot with a magnetic tactile sensor mounted on the back of the hand.
}
    \caption{}
  \end{subfigure}
  \caption{(a) Test setup for robotic grasping (Dofbot-pi4B). Note that the magnetic tactile sensor was mounted on the inner side of the gripper. (b) Test setup for humanoid robot (Unitree G1). Note that the magnetic tactile sensor was mounted on the back of the hand.}
  \label{fig:evaluation1}
\end{figure}

\subsection{Fabrication of Stealth Tags}
Utilizing the reconstructed response functions (as we presented in Sec. ~\ref{sec:design1}), we localized optimal coordinates for magnetic perturbation. 
We adopted a vertical offset of $Z=0.2$~mm as our design baseline, corresponding to the maximum thickness of standard commercial labels (Table~\ref{tab:label_thickness}).

First, we utilize the constructed continuous spatial response functions to perform an analysis at the $Z=0.2~\mathrm{mm}$ plane. 
{\name} needs to ensure decoupled signal injection.
That is, if the attacker intends to inject a normal force, then an effective attack should minimize the injected force in the X-Y plane. To achieve this goal, we prioritized ``independent regions'' where the sensitivity of the target force component is maximized while interference with other force components is minimized, e.g., the regions highlighted in Fig.~\ref{fig:design2}.

After determining the relatively independent sensitive regions, we place a magnet with a diameter of $3~\mathrm{mm}$ at the center of this region to serve as the main attack magnet.
It should be noted that the magnetic tactile sensor typically has multiple sensitive regions, and the directions of the interpreted force in different regions may be opposite.
This is because when constructing the dataset (as shown in Sec.~\ref{sec:vulnerability}), the magnet used was oriented with the N-pole facing the magnetic tactile sensor. 
Therefore, the response distribution in Fig.~\ref{fig:design2} corresponds to the scenario where the N-pole faces the sensor. 
Based on our empirical tests, when the S-pole faces the sensor, the direction of the force response will be reversed. 
For example, when the N-pole faces the sensor, a response of $F_x=1.5~\mathrm{N}$ is generated at a certain location. 
If the same magnet is changed to have the S-pole facing the sensor at the same location, a response of $F_x=-1.5~\mathrm{N}$ will be induced. 
Therefore, we can combine the spatial distribution of the response regions and the polarity direction of the magnets to deploy multiple main attack magnets in order to enhance the injection effect on the target force component.

Furthermore, in real-world application scenarios, certain offsets may exist in robotic arm grasping, robot motion, and the contact position between the tag and the sensor. 
To improve the fault tolerance and stability of the attack, we deploy auxiliary attack magnets with a diameter of $2.5~\mathrm{mm}$ around the main attack magnet based on the distribution characteristics of the sensitive regions.
We present the tag layout in Fig.~\ref{fig:design2}. 
We evaluated the performance of this design and presented the results in Sec.~\ref{sec:Robustness}. Beyond the single-tag design, multiple {\name} tags could be deployed to cover a wider range of likely contact regions. 
For example, multiple tags could be arranged as a distributed array over likely contact areas on large objects. 
For symmetric objects, one can deploy tags by following a circumferential layout around the contact regions, e.g., the centroid of the object. 
Such geometry-aware placements can enlarge the effective contact region and improve tolerance to contact-position variations.

Finally, we implement the designed magnet layout by placing magnets on a grid paper. 
We stabilize each magnet at its predetermined position with adhesives.

\section{Evaluation}

\subsection{Evaluation Setup}

To highlight the end-to-end impact of {\name}, we conducted real-world experiments on two robotic platforms. 
We used a Dofbot-Pi4B~\cite{yahboom_dofbot_pi} robotic arm to assess {\name}'s performance in a popular grasping task.
As shown in Fig.~\ref{fig:evaluation1}(a), the robot arm's goal is to move a small carton. 
With a tactile sensor embedded on the claw, one can achieve a closed-loop force control, thus assuring the gripper will not drop or crush the object.
We also used a Unitree G1~\cite{unitree_g1_developer} humanoid robot to evaluate a collision detection scenario.
As shown in Fig.~\ref{fig:evaluation1}(b), the on-body tactile sensor should inform the robot when a contact occurs, thus preventing it from breaking a fragile object, e.g., a vase.
Note that this control loop allows the robot to perform safe manipulation without seeing the object, mimicking how humans would perform safe tasks.
In both experiments, {\name}'s tag was concealed by a camouflage (i.e., a QR code), which can be replaced by other camouflage labels as shown in Fig.~\ref{fig:vulnerability4}.
We will elaborate on the details of these experiments in Sec.~\ref{sec:robotic_grasping} and Sec.~\ref{subsec:humanoid}, respectively.

For magnetic tactile sensors, we adopted PaXini PX6AX-GEN3-MC-M2020-Elite.  As shown in Sec.~\ref{sec:background2.3}, PaXini shows stronger resistance to ambient magnetic fields incurred by everyday electronics compared with other major models (e.g., XELA uSPa 46), thus demonstrating superb performance against magnetic interference.
We therefore selected the PaXini model as the primary experimental platform.

\begin{figure*}[t]
    \centering
    \begin{subfigure}[b]{0.495\textwidth}
        \centering
        \includegraphics[width=\textwidth]{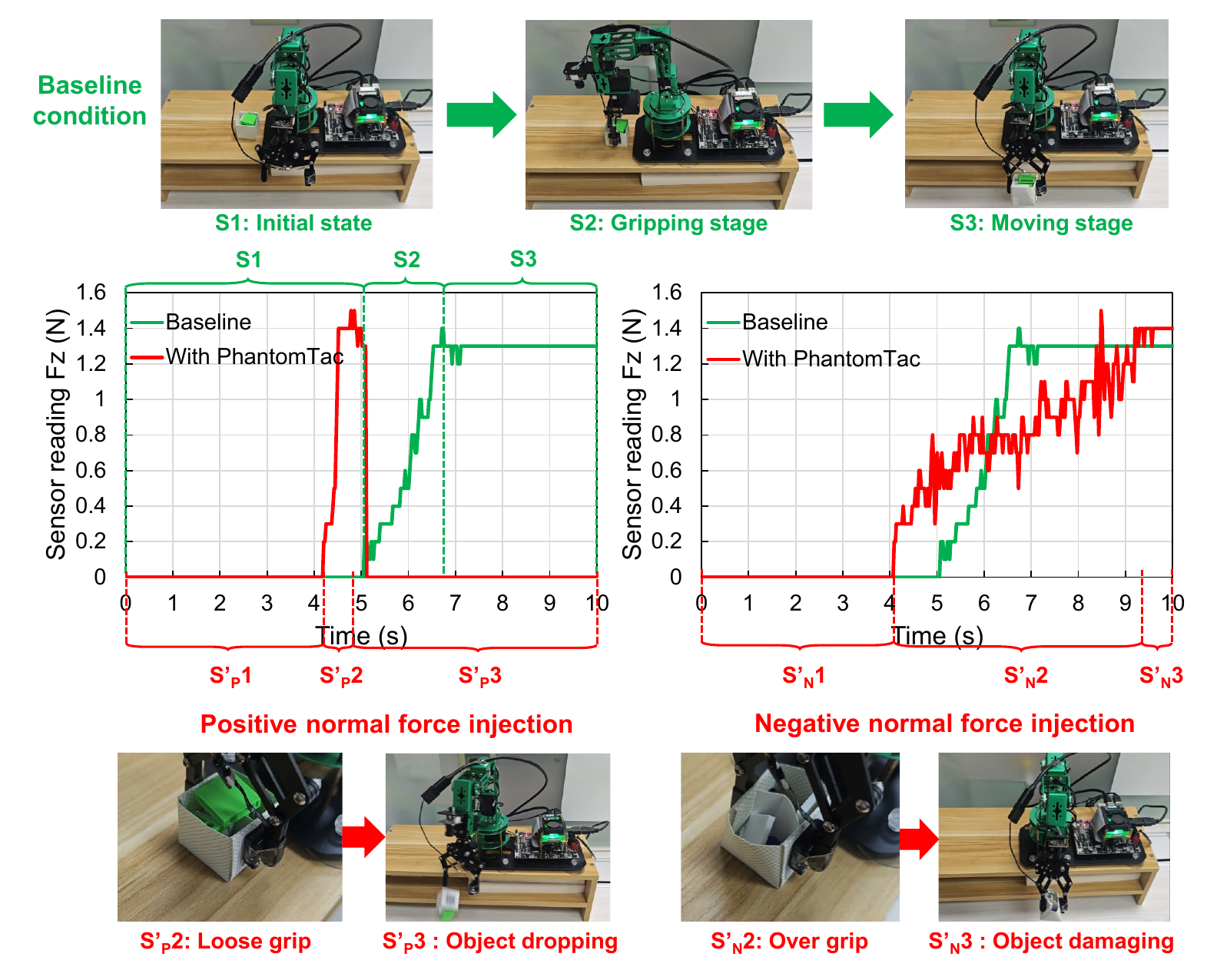}
        \Description{
        End-to-end attack results of PhantomTac on a robotic manipulator, showing positive and negative normal force injection during the grasping task across different stages of operation.
        }
        \caption{{\name}'s attack on a robotic arm.}
        \label{fig:subfig1}
    \end{subfigure}
    \hfill
    \begin{subfigure}[b]{0.5\textwidth}
        \centering
        \includegraphics[width=\textwidth]{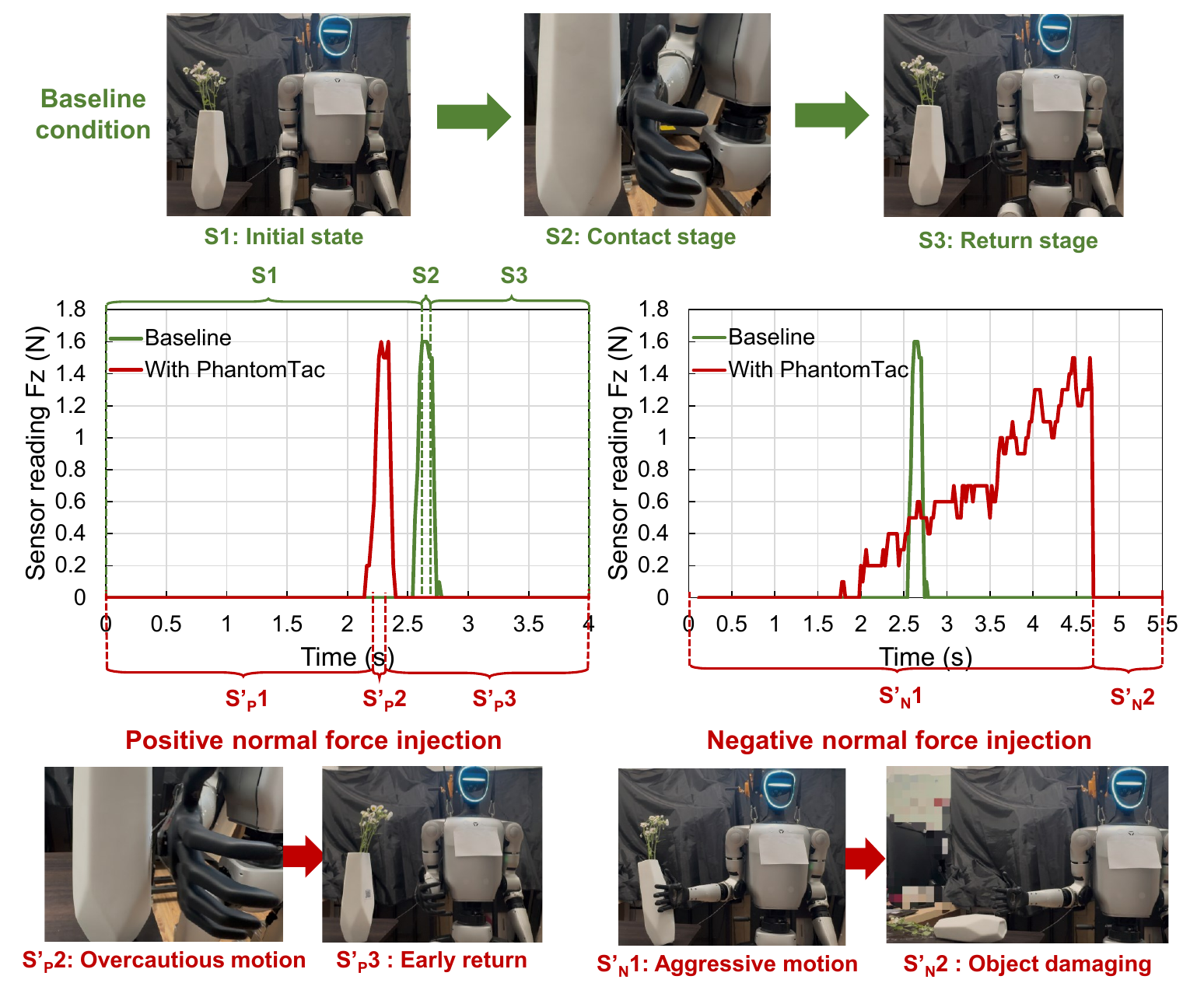}
        \Description{
        End-to-end attack results of PhantomTac on a humanoid robot, showing positive and negative normal force injection during the collision detection task across different stages of operation.
        }
        \caption{{\name}'s attack on a humanoid robot. }
        \label{fig:subfig2}
    \end{subfigure}
    \caption{End-to-end attack results on the robotic manipulator and the humanoid robot. 
    (a) Positive and negative normal force injection in the grasping task. 
    (b) Positive and negative normal force injection in the collision detection task.
    Note that S1, S2, and S3 denote the different stages of a benign operation.
    $S_{p}^{\prime}1$, $S_{p}^{\prime}2$, and $S_{p}^{\prime}3$ denote stages under positive force injection, whereas $S_{n}^{\prime}1$, $S_{n}^{\prime}2$, and $S_{n}^{\prime}3$ present negative ones.
    }
    \label{fig:mainfig}
\end{figure*}

\subsection{End-to-End Evaluation \#1: Robotic Grasping}
\label{sec:robotic_grasping}
The robotic arm followed a predetermined trajectory and executed a tactile closed-loop feedback controller adapted from existing open-source solutions, i.e., LeTac-MPC~\cite{xu2024letacmpc}. 
Specifically, the controller monitored normal force ($F_z$) at high frequency and regulated gripper motion through force hysteresis: it tightened when $F_z$ drops below a preset lower limit, maintained its pose when $F_z$ is within a safe range, and loosened when $F_z$ exceeds the upper limit. 

In a benign operation, the robotic arm grasped a small paper box, with a grasping-force threshold of 1.3--2~N. The normal grasping task typically contains three stages: approach (i.e., stage $S1$ as shown in Fig.~\ref{fig:mainfig}), grasp and stop at force threshold ($S2$), and transport ($S3$). During normal operation, the arm moves to a predefined pose, closes the gripper, and stops once tactile feedback reaches the threshold. The system then collects 5 tactile feedback frames (approximately 0.1~s at 50~Hz) to assess grasp stability. If all 5 consecutive force readings remain within the threshold range, the arm initiates transport as shown in Fig.~\ref{fig:mainfig}(a).
To establish a baseline, we performed 20 grasping trials in an interference-free environment, and they all succeeded without damaging the box, confirming the controller's reliability in normal operation.

\subsubsection{Positive Normal-Force Injection}
Now we assess {\name}'s efficacy in this task. Specifically, positive injection aims to increase the perceived force of the sensor.
For example, when the actual force is 1 N and a positive normal force of 1 N is injected, the normal force interpreted by the tactile sensor will rise to 2 N.

Without attack, the robotic arm contacted the object at approximately 5~s and reached the force threshold at 6.5~s. The system then assesses the grasp stability and initiates transport.
With positive normal-force injection, the sensor prematurely generates force readings at 4.1~s ($S'_P2$) due to the injected magnetic field. As the gripper closes, the injected force increases rapidly (magnetic field strength decays with distance cubed). The sensor reaches the force threshold at 4.5~s($S'_P3$), triggering premature stability assessment. The arm incorrectly determines the grasp is stable and begins transport. However, since the gripper has not yet firmly contacted the object, insufficient grasping force causes the object to drop during transport.

\subsubsection{Negative Normal-Force Injection}
Specifically, negative injection decreases the perceived force (e.g., based on an initial perceived force of 2~N, after injecting a negative normal force of 1~N, the normal force perceived by the tactile sensor will drop to 1~N).

With a negative injection of normal force, the sensor begins generating force readings at 4~s ($S'_N2$) but rises slowly because the injected signal cancels out a portion of the actual contact feedback. 
As the gripper continues closing, the object undergoes progressive deformation and the sensor readings exhibit significant fluctuations. The sensor does not reach the force threshold until approximately 9~s ($S'_N3$). By this point, the object has already experienced significant deformation from excessive grasping force. Because the negative normal force injection delayed the timing for the gripper to stop closing, the robotic arm continuously applied an excessive gripping force, ultimately causing the object to be excessively squeezed and damaged.

\subsection{End-to-End Evaluation \#2: Humanoid Motion}
\label{subsec:humanoid}
For collision detection, the humanoid robot followed a fixed trajectory and executed a collision detection algorithm adapted from Danfoa~\cite{danfoaForceSensorUtils} for the robotic arm-environment interaction. 
The collision force threshold was set to 1.5 --- 2.0~N, allowing stable contact with slight wobbling without knocking the vase over. During normal operation, the robot extends its arm along a predefined trajectory ($S1$). The system collects five tactile feedback frames (approximately 0.1~s at 50~Hz) to assess collision validity ($S2$). If all 5 consecutive force readings remain within the threshold range, the robot determines an effective collision has occurred and initiates retraction ($S3$). 

In the experiment, the algorithm drove the robotic arm toward the target area at a constant low speed while continuously sampling normal force from local tactile sensors. 
If the normal force exceeds the predefined contact threshold for several consecutive frames, the system will report a collision, trigger an emergency stop, and then return to its original position along the original path. Otherwise, it continues to execute the predefined waypoints. 
We also performed 20 collision detection trials in an interference-free environment. 
All trials correctly detected the contact, triggered the stop, and returned to the original position along the original path, verifying the reliability of the algorithm.

\subsubsection{Positive Normal-Force Injection}
As depicted in Fig.~\ref{fig:mainfig}, without attack, the robot contacts the vase at approximately 2.5~s and reaches the force threshold at 2.7~s. 
The system assesses collision validity and initiates retraction.
With positive normal-force injection, the sensor prematurely generates force readings at 2.1~s ($S'_P1$). The injected force increases as the robot moves, reaching the threshold at 2.3~s ($S'_P2$) and triggering premature collision detection ($S'_P3$). The robot incorrectly determines an effective collision has occurred and prematurely initiates retraction. Since the vase has not yet been contacted, this demonstrates that {\name} can generate false contact feedback and disrupt normal task execution (i.e., stalling as shown in Fig.~\ref{fig:overview of PhantomTac}).

\subsubsection{Negative Normal-Force Injection}
With negative normal-force injection, the sensor begins generating force readings at 2~s ($S'_N1$) but rises slowly because the injected signal cancels out a portion of the actual contact feedback. 
As the robot continues moving, the pushing force on the vase gradually increases, causing progressive tilting and sensor fluctuations. 
At approximately 4.7~s ($S'_N2$), the vase toppled over due to excessive tilting, losing contact with the sensor, and the sensor readings consequently returned to zero. 
Because the negative normal force injection delays the timing for the robot's motion to stop, the robot applied an excessive pushing force, ultimately causing the vase to be pushed over and damaged.

\subsection{Robustness to Grasp Position Offset}
\label{sec:Robustness}
\begin{figure*}[t] 
  \centering
  \includegraphics[width=0.95\textwidth]{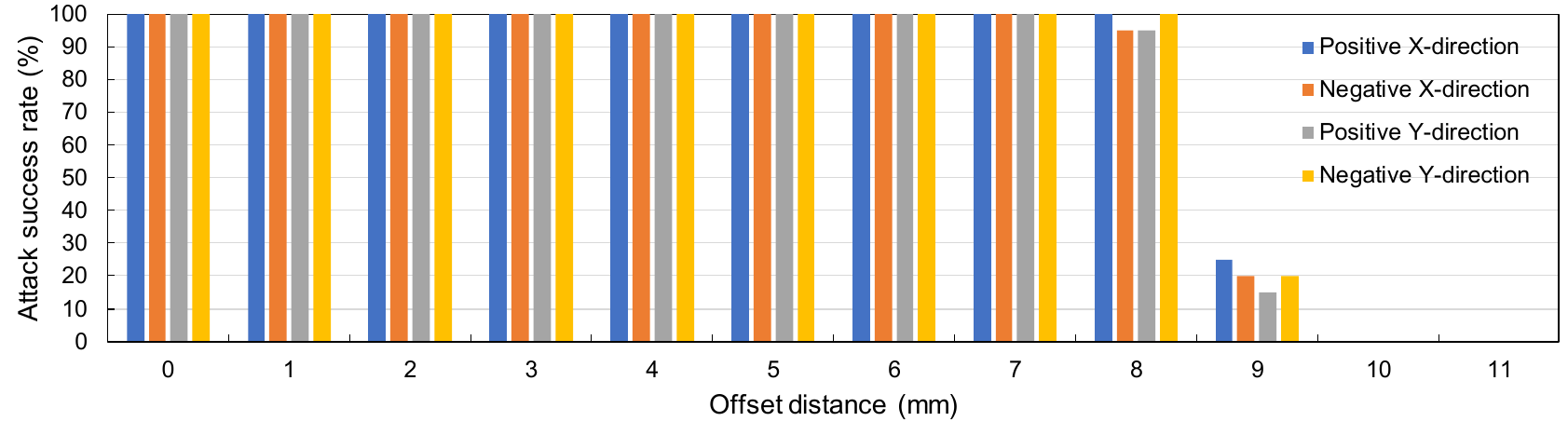} 
  \Description{Robustness evaluation results of PhantomTac under contact position variation, showing how changes in the contact location affect the attack performance of the magnetic tactile sensing system.}
  \caption{Result of robustness test for contact position variation.}
  \label{fig:evaluation2}
\end{figure*}
In this experiment, we assess {\name}'s performance under positional variations.
Modern robotic systems are capable of performing high-accuracy tasks repetitively.
For example, the positioning repeatability of the grasping system of the FANUC LR Mate 200iD can reach 0.01~mm \cite{fanucLRMate200iD}.
The gripper's pose repeatability of Universal Robots UR5e is 0.03~mm \cite{ur5eSpec}. 
In addition, Kulecki et al.~\cite{kulecki2021practical} evaluated an object detection and grasping system for a mobile manipulator in unstructured environments. 
In their experiment, the average position error remained below 6 mm. 
However, it is still possible that the position of each grasp can vary.

To evaluate the robustness of the designed magnetic tags against contact-position variations, we used the robotic arm setup described in Sec.~\ref{sec:robotic_grasping}, and defined the X and Y axes of the sensor's contact interface as shown in Fig.~\ref{fig:evaluation1}. 
We implemented the attack by injecting a positive normal-force signal, and progressively shifted the grasping position during injection. 
This resulted in misalignment between the gripper and {\name} when contacting the target object. 
We considered four offset directions: the positive and negative directions along the X-axis and Y-axis. 
For each direction, we increased the grasping-position offset from 0 mm to 11 mm in 1 mm increments. 
At each offset distance, we performed 20 grasping trials and computed the attack success rate.

The results are shown in Fig.~\ref{fig:evaluation2}. When the offset along any of the four directions is within 8~mm, the attack success rate remains above 95\%. 
The attack success rate drops when the offset reaches 9~mm. 
These results show that {\name} can tolerate moderate grasping-position offsets (up to 8~mm) in our experimental setup.

\subsection{Generalizability of {\name} Across Different Models}
\label{sec:across_models}
\begin{figure}[t] 
  \centering
  \includegraphics[width=0.92\linewidth]{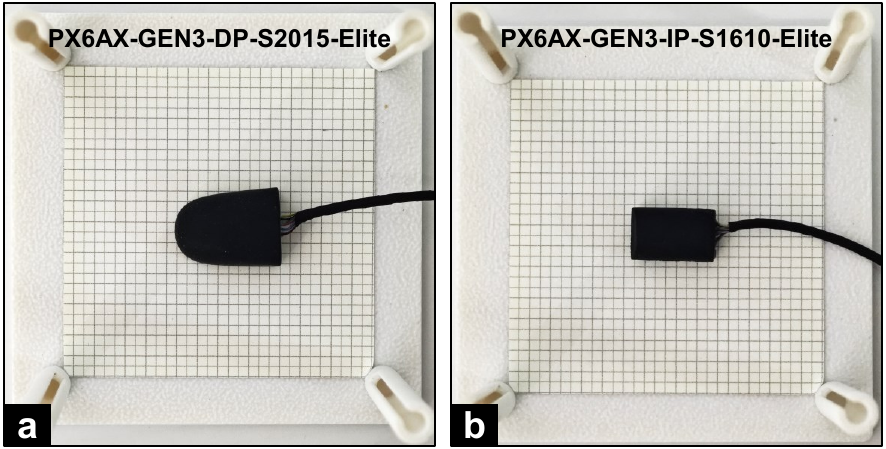}
  \Description{Experimental setup for evaluating the generalizability of PhantomTac across different magnetic tactile sensors, showing two sensor models used in the evaluation: PX6AX-GEN3-DP-S2015-Elite and PX6AX-GEN3-IP-S1610-Elite.}
  \caption{Generalizability experiment setup. 
  (a) PX6AX-GEN3-DP-S2015-Elite. (b) PX6AX-GEN3-IP-S1610-Elite.
  }
  \label{fig:evaluation4}
\end{figure}
In previous experiments (Sec.~\ref{sec:background}), we compared two commercial magnetic tactile sensors: PaXini PX6AX-GEN3-MC-M2020-Elite and XELA uSPa 46.
To verify the generalizability of the proposed vulnerability, we conducted additional tests on two other PaXini sensors: the finger-pad PX6AX-GEN3-IP-S1610-Elite and the fingertip PX6AX-GEN3-DP-S2015-Elite. 
We present their form factors and experiment setup on grid papers in Fig.~\ref{fig:evaluation4}.

Specifically, under the same experimental setup as in Sec.~\ref{sec:background}, we conducted magnetic interference resistance experiments on the finger pad PX6AX-GEN3-IP-S1610-Elite and the fingertip PX6AX-GEN3-DP-S2015-Elite.
For electronics, none of these devices can interfere with the finger pad and fingertip tactile sensors.
For the passive magnet, the empirical results indicate that the fingertip model continuously outputs a force with a vector sum of 1.31~N. 
This value is only $85.6\%$ of the response from the palm model (i.e., PX6AX-GEN3-MC-M2020-Elite), which outputs 1.53~N.
The finger-pad model continuously outputs a force with a vector sum of 1.05~N. 
This value is approximately $68.6\%$ of the response from the palm one.
Although the finger pad and fingertip models demonstrated better resilience than the palm model, a focused magnetic field (e.g., by a small magnet) can still incur distinctive interferences --- much stronger than the interference incurred by common electronic devices.
{\emergencystretch=1em
These results show that the underlying vulnerability generalizes across the evaluated magnetic tactile sensor models and is not specific to a single manufacturer or model. 

\par}

\section{Countermeasures and Discussion}
\label{sec:Countermeasures}

\subsection{Countermeasures}
The core attack of {\name} is that an attacker can alter the output of the magnetic tactile sensor via an external magnetic tag.
To address this issue, we discuss a potential defense strategy grounded in multimodal sensor fusion.

Modern robotic or manipulator systems typically provide additional motion-related signals, such as joint torques, gripper motor currents ~\cite{frankaRobotState}, and the pose, velocity ~\cite{urRTDEGuide}, and motion status of the end-effector ~\cite{robotiq2fManual}. 
Although these signals may not provide the same spatial resolution as tactile sensors, they offer independent sources of physical interaction information. 
Such information can be used to verify whether the tactile response is consistent with the actual motion of the robot.

Specifically, the controller can maintain a short time window for both tactile readings and proprioceptive signals. 
A contact event is accepted only when the increase in tactile force is accompanied by other varying signals, such as an increase in gripper motor current, an increase in joint torque, a decrease in gripper closing velocity, or restricted movement of the end-effector.
If the tactile sensor reports a large normal force while the gripper motor current remains close to the neutral pose, the system can flag this as a potential false contact event and prevent the controller from prematurely stopping the gripper.

Similarly, during a negative normal force injection attack, the tactile sensor may continue to report insufficient contact force even if the gripper has already contacted the object. 
In this case, the system can compare the tactile signal with the gripper motor load, joint torques, and gripper motion status. 
If the gripper motor current or joint torque increases while the gripper closing velocity decreases or the end-effector movement is mechanically restricted, the controller can infer that physical contact has actually occurred despite the low tactile readings. 
Then, the system can stop further increases in force or switch to a safe recovery strategy, rather than continuing to apply pressure on the object.

In practical applications, this defense can be implemented as a lightweight monitoring module. 
For each control cycle, it checks whether the tactile force and other signals show a consistent pattern within a sliding window. 
When inconsistencies persist over several consecutive frames, the robotic system can trigger an alarm, ignore suspicious tactile readings, or enter a conservative fail-safe mode. 
This design aims to reduce false positives while making tactile spoofing attacks more difficult.

\subsection{
Discussion
}

\noindent\textbf{Cross-model applicability.}
As shown in Sec.~\ref {sec:background_limitation} and Sec.~\ref {sec:across_models}, the underlying vulnerability is present across multiple evaluated magnetic tactile sensor models. However, the sensor responses differ across models. Therefore, deterministic force injection remains model-specific and may require offline characterization of the target model and corresponding tag customization. As described in Sec.~\ref {sec:reverse engineering}, the offline characterization can be performed on a separately obtained sensor of the same model and requires no access to the robotic system.

{\emergencystretch=1em
\noindent\textbf{Sensing-modality scope.}
This cross-model generalizability is modality-specific. {\name} relies on magnetic-field manipulation and therefore directly targets magnetic sensing components. It does not directly apply to purely optical, capacitive, or resistive tactile sensors. In multimodal tactile systems, {\name} can affect the magnetic sensing component, while its system-level impact depends on the sensor configuration and fusion mechanism.

\par}

\section{Related Work}

\noindent\textbf{Tactile Sensing Technologies.} Tactile sensors are capable of direct, contact-level feedback on force, slip, and material properties~\cite{dahiya2009tactile, li2020tactile}. 
Recently, a diverse set of transduction principles has been extensively explored.
Vision-based tactile sensors, e.g., GelSight~\cite{yuan2017gelsight}, DIGIT~\cite{lambeta2020digit}, and TacTip~\cite{ward2018tactip}, capture elastomer deformation with an embedded camera and extract high-resolution geometry and force fields through computer-vision pipelines.
However, they are typically constrained by camera frame rate, image-processing overhead, and bulky optical packaging.
Capacitive tactile sensors measure deformation-induced capacitance changes~\cite{schmitz2011methods}.
However, this approach falls short in sensing fine-grained force information and is susceptible to parasitic capacitance, humidity, and temperature drift.
Piezoresistive tactile sensors exploit force-induced resistance changes~\cite{stassi2014flexible}, thus having lower costs.
However, this approach suffers from hysteresis, drift, and limited long-term stability.
In contrast, magnetic tactile sensors transduce contact forces by tracking the displacement of an embedded magnet with a Hall-effect or magnetoresistive array~\cite{bhirangi2021reskin, yan2021softmagnetic}, enabling fine-grained three-dimensional force decoupling, compact form factors, robustness to optical occlusion, and low fabrication cost.

\noindent\textbf{Physical-Layer Signal Injection Attacks.} 
A broad class of prior work has demonstrated that adversaries can manipulate embedded sensors by injecting crafted physical signals, bypassing higher-level software defenses.
Kune et al.~\cite{kune2013ghost} first systematically characterized electromagnetic signal injection against analog sensors, showing that EMI can induce spurious readings in implantable cardiac devices and consumer microphones without any physical contact.
Trippel et al.~\cite{trippel2017walnut} demonstrated that MEMS accelerometers in wearables and drones can be driven to arbitrary output values via acoustic resonance injection, corrupting the inertial measurements that autonomous systems depend on.
Ji et al.~\cite{ji2021poltergeist} extended this idea to vision pipelines, using acoustic signals to cause resonance-induced blur in camera image sensors, thereby degrading the perception of computer-vision-based autonomous vehicles.
Xu et al.~\cite{xu2018analyzing} analyzed signal injection vulnerabilities in ultrasonic sensors used for obstacle detection in autonomous vehicles, demonstrating that spoofed echo signals can suppress or fabricate obstacles.
Zhang et al.~\cite{zhang2017dolphinattack} and Yan et al.~\cite{yan2019feasibility} showed that voice assistants are vulnerable to inaudible ultrasonic command injection, which is imperceptible to humans yet recognized as valid speech by microphone front-ends.
{\name} shares this adversarial philosophy but targets a qualitatively different modality---magnetic tactile sensing---and introduces a uniquely passive, \emph{deploy-and-forget} attack vector using battery-free permanent magnet tags, whereas prior injection attacks typically require active electronic signal generators in close proximity.

\noindent\textbf{Security of Robotic Perception and Tactile Sensing.}
As embodied AI systems are expanding to safety-critical settings, the security of robotic perception has attracted increasing attention. 
Prior work has examined adversarial attacks on visual pipelines~\cite{ji2021poltergeist}, ultrasonic obstacle sensors~\cite{xu2018analyzing}, and inertial measurement units~\cite{trippel2017walnut}.
However, limited efforts have been made into the potential security issues of tactile sensing.
Existing anti-interference mechanisms in magnetic tactile sensors, such as gradiometric differential arrays~\cite{le2022gradiometric}, are only capable of overcoming common interfering magnetic fields. 
To the best of our knowledge, {\name} is the first effort to characterize and exploit the security loophole in magnetic tactile sensors. 
{\name} demonstrates the end-to-end hijacking performance of robotic closed-loop force control with an attack success rate exceeding 95\%.

\section{Conclusion}
\label{sec:conclusion}
Magnetic tactile sensors have been emerging as major approaches for achieving fine-grained force sensing. 
{\name} revealed a physical loophole of magnetic tactile sensors. That is, they are susceptible to magnetic injection. 
This leads to normal and shear forces injection, thus undermining the safety of robotic systems.
By performing a novel reverse engineering effort, we can fabricate the deploy-and-forget, battery-free tags, which can be as thin as a piece of paper.
We verified {\name}'s efficacy in end-to-end systems on both robotic arm and humanoid robots.
We believe {\name} opens research directions for enabling more reliable tactile sensing technologies.

\begin{acks}
We would like to thank the anonymous reviewers for their helpful
comments. This work was supported by the National Natural Science Foundation of China under Grant No. 62472283.

\end{acks}

\bibliographystyle{ACM-Reference-Format}
\bibliography{bib}


\begin{thebibliography}{59}


\ifx \showCODEN    \undefined \def \showCODEN     #1{\unskip}     \fi
\ifx \showISBNx    \undefined \def \showISBNx     #1{\unskip}     \fi
\ifx \showISBNxiii \undefined \def \showISBNxiii  #1{\unskip}     \fi
\ifx \showISSN     \undefined \def \showISSN      #1{\unskip}     \fi
\ifx \showLCCN     \undefined \def \showLCCN      #1{\unskip}     \fi
\ifx \shownote     \undefined \def \shownote      #1{#1}          \fi
\ifx \showarticletitle \undefined \def \showarticletitle #1{#1}   \fi
\ifx \showURL      \undefined \def \showURL       {\relax}        \fi
\providecommand\bibfield[2]{#2}
\providecommand\bibinfo[2]{#2}
\providecommand\natexlab[1]{#1}
\providecommand\showeprint[2][]{arXiv:#2}
\makeatletter
\@ifundefined{NAT@parse@date}{}{\let\NAT@parse@date@orig\NAT@parse@date}
\@ifundefined{NAT@parse@date}{}{\def\NAT@parse@date#1#2#3#4#5#6@@{\NAT@parse@date@orig#1#2#3#4#5#6@@\def\NAT@tempyear{0000}\def\NAT@tempexlab{{?}}\ifx\NAT@year\NAT@tempyear\ifx\NAT@exlab\NAT@tempexlab\def\NAT@date{[n.\,d.]}\else\edef\NAT@date{[n.\,d.]\NAT@exlab}\fi\fi}}
\makeatother

\bibitem[Andrussow et~al\mbox{.}(2023)]%
        {andrussow2023minsight}
\bibfield{author}{\bibinfo{person}{Iris Andrussow}, \bibinfo{person}{Huanbo
  Sun}, \bibinfo{person}{Katherine~J. Kuchenbecker}, {and}
  \bibinfo{person}{Georg Martius}.} \bibinfo{year}{2023}\natexlab{}.
\newblock \showarticletitle{Minsight: A Fingertip-Sized Vision-Based Tactile
  Sensor for Robotic Manipulation}.
\newblock \bibinfo{journal}{\emph{Advanced Intelligent Systems}}
  \bibinfo{volume}{5}, \bibinfo{number}{8}, Article
  \bibinfo{articleno}{2300042} (\bibinfo{year}{2023}).
\newblock
\href{https://doi.org/10.1002/aisy.202300042}{doi:\nolinkurl{10.1002/aisy.202300042}}


\bibitem[{Avery Dennison}(2010)]%
        {avery_synthetic_paper_bw0153}
\bibfield{author}{\bibinfo{person}{{Avery Dennison}}.}
  \bibinfo{year}{2010}\natexlab{}.
\newblock \bibinfo{booktitle}{\emph{{Fasson Synthetic Paper/S4700N/BG40 Wht
  (Imp.) Product Data Sheet}}}.
\newblock Avery Dennison.
\newblock
\urldef\tempurl%
\url{https://label.averydennison.asia/content/dam/averydennison/lpm/ap/asean/product-families/Data-Sheets/Film/BW0153%2C%20Fasson-%20Synthetic%20Paper_S4700N_B.pdf}
\showURL{%
\tempurl}
\newblock
\shownote{Spec. BW0153. Last modified: 18 Mar 2010. Accessed: 2026-04-28}.


\bibitem[{Avery Dennison}(2021)]%
        {avery_direct_thermal_200gp}
\bibfield{author}{\bibinfo{person}{{Avery Dennison}}.}
  \bibinfo{year}{2021}\natexlab{}.
\newblock \bibinfo{booktitle}{\emph{{Fasson Direct Thermal 200GP
  FSC/S1002/BG33Wh FSC Product Data Sheet}}}.
\newblock Avery Dennison Label and Packaging Materials.
\newblock
\urldef\tempurl%
\url{https://label.averydennison.com/content/dam/averydennison/lpm-responsive/asia-pacific/en-sa/documents/product-data-sheets/asean/NW0060F-pds-en.pdf}
\showURL{%
\tempurl}
\newblock
\shownote{Product code NW0060F. Updated 01 Sept 2021. Accessed: 2026-04-28}.


\bibitem[{Avery Dennison}(2022)]%
        {avery_pet_sy7201}
\bibfield{author}{\bibinfo{person}{{Avery Dennison}}.}
  \bibinfo{year}{2022}\natexlab{}.
\newblock \bibinfo{booktitle}{\emph{{Fasson 2M WH PET TC/S692N/BG40Wh Imp
  Product Data Sheet}}}.
\newblock Avery Dennison Label and Packaging Materials.
\newblock
\urldef\tempurl%
\url{https://label.averydennison.com/content/dam/averydennison/lpm-responsive/asia-pacific/en-sa/documents/product-data-sheets/asean/SY7201-pds-en.pdf}
\showURL{%
\tempurl}
\newblock
\shownote{Product code SY7201. Updated 07 Jun 2022. Accessed: 2026-04-28}.


\bibitem[{Avery Dennison}(2024)]%
        {avery_high_gloss_paper}
\bibfield{author}{\bibinfo{person}{{Avery Dennison}}.}
  \bibinfo{year}{2024}\natexlab{}.
\newblock \bibinfo{booktitle}{\emph{{Fasson High Gloss Paper/S1010/rCCK80Wh Imp
  Product Data Sheet}}}.
\newblock Avery Dennison Label and Packaging Materials.
\newblock
\urldef\tempurl%
\url{https://label.averydennison.com/content/dam/averydennison/lpm-responsive/asia-pacific/en-sa/documents/product-data-sheets/asean/AW0335-pds-en.pdf}
\showURL{%
\tempurl}
\newblock
\shownote{Product code AW0335. Updated 09 Sept 2024. Accessed: 2026-04-28}.


\bibitem[Bhirangi et~al\mbox{.}(2025)]%
        {bhirangi2025anyskin}
\bibfield{author}{\bibinfo{person}{Raunaq Bhirangi}, \bibinfo{person}{Venkatesh
  Pattabiraman}, \bibinfo{person}{Enes Erciyes}, \bibinfo{person}{Yifeng Cao},
  \bibinfo{person}{Tess Hellebrekers}, {and} \bibinfo{person}{Lerrel Pinto}.}
  \bibinfo{year}{2025}\natexlab{}.
\newblock \showarticletitle{AnySkin: Plug-and-Play Skin Sensing for Robotic
  Touch}. In \bibinfo{booktitle}{\emph{Proceedings of the 2025 IEEE
  International Conference on Robotics and Automation}}.
  \bibinfo{pages}{16563--16570}.
\newblock
\href{https://doi.org/10.1109/ICRA55743.2025.11128638}{doi:\nolinkurl{10.1109/ICRA55743.2025.11128638}}


\bibitem[Bhirangi et~al\mbox{.}(2021)]%
        {bhirangi2021reskin}
\bibfield{author}{\bibinfo{person}{Raunaq~M. Bhirangi},
  \bibinfo{person}{Tess~Lee Hellebrekers}, \bibinfo{person}{Carmel Majidi},
  {and} \bibinfo{person}{Abhinav Gupta}.} \bibinfo{year}{2021}\natexlab{}.
\newblock \showarticletitle{ReSkin: versatile, replaceable, lasting tactile
  skins}. In \bibinfo{booktitle}{\emph{Conference on Robot Learning, 8-11
  November 2021, London, {UK}}} \emph{(\bibinfo{series}{Proceedings of Machine
  Learning Research})}, \bibfield{editor}{\bibinfo{person}{Aleksandra Faust},
  \bibinfo{person}{David Hsu}, {and} \bibinfo{person}{Gerhard Neumann}} (Eds.).
  \bibinfo{publisher}{{PMLR}}, \bibinfo{pages}{587--597}.
\newblock
\urldef\tempurl%
\url{https://proceedings.mlr.press/v164/bhirangi22a.html}
\showURL{%
\tempurl}


\bibitem[Dahiya et~al\mbox{.}(2010)]%
        {dahiya2009tactile}
\bibfield{author}{\bibinfo{person}{Ravinder~S. Dahiya},
  \bibinfo{person}{Giorgio Metta}, \bibinfo{person}{Maurizio Valle}, {and}
  \bibinfo{person}{Giulio Sandini}.} \bibinfo{year}{2010}\natexlab{}.
\newblock \showarticletitle{Tactile Sensing - From Humans to Humanoids}.
\newblock \bibinfo{journal}{\emph{{IEEE} Trans. Robotics}}
  \bibinfo{volume}{26}, \bibinfo{number}{1} (\bibinfo{year}{2010}),
  \bibinfo{pages}{1--20}.
\newblock
\href{https://doi.org/10.1109/TRO.2009.2033627}{doi:\nolinkurl{10.1109/TRO.2009.2033627}}


\bibitem[{Danfoa}(2019)]%
        {danfoaForceSensorUtils}
\bibfield{author}{\bibinfo{person}{{Danfoa}}.} \bibinfo{year}{2019}\natexlab{}.
\newblock \bibinfo{title}{{Force Sensor Utils: ROS Package for Force/Torque
  Sensor Processing and Monitoring}}.
\newblock
  \bibinfo{howpublished}{\url{https://github.com/Danfoa/force_sensor_utils}}.
\newblock
\newblock
\shownote{GitHub repository. Accessed: 2026-04-30}.


\bibitem[{Ebuy7}(2026)]%
        {ebuy7_nscing_889101}
\bibfield{author}{\bibinfo{person}{{Ebuy7}}.} \bibinfo{year}{2026}\natexlab{}.
\newblock \bibinfo{title}{{Nscing ES 889-101 Digital Height Gauge (0--300/500
  mm, Stainless Steel)}}.
\newblock
  \bibinfo{howpublished}{https://www.ebuy7.com/ms/suce-high-precision-height-ruler-digital-display-cursor-0-300-electronic-caliper-0-500mm-line-and.html}.
\newblock
\newblock
\shownote{Product page for Nscing es 889-101 height gauge. Accessed:
  2026-04-28}.


\bibitem[{FANUC America}(2026)]%
        {fanucLRMate200iD}
\bibfield{author}{\bibinfo{person}{{FANUC America}}.}
  \bibinfo{year}{2026}\natexlab{}.
\newblock \bibinfo{title}{{LR Mate 200iD Robot - Small Payload Robot}}.
\newblock
\urldef\tempurl%
\url{https://www.fanucamerica.com/products/robots/series/lr-mate/lr-mate-200id}
\showURL{%
\tempurl}
\newblock
\shownote{Accessed: 2026-04-30}.


\bibitem[{Franka Robotics}(2026)]%
        {frankaRobotState}
\bibfield{author}{\bibinfo{person}{{Franka Robotics}}.}
  \bibinfo{year}{2026}\natexlab{}.
\newblock \bibinfo{title}{{libfranka}: RobotState Struct Reference}.
\newblock
\urldef\tempurl%
\url{https://frankarobotics.github.io/libfranka/0.15.0/structfranka_1_1RobotState.html}
\showURL{%
\tempurl}
\newblock
\shownote{Accessed: 2026-04-30}.


\bibitem[{GelSight}(2026a)]%
        {gelsight2026official}
\bibfield{author}{\bibinfo{person}{{GelSight}}.}
  \bibinfo{year}{2026}\natexlab{a}.
\newblock \bibinfo{title}{{3D} Tactile Sensing with {GelSight}'s Elastomeric
  Platform}.
\newblock
\urldef\tempurl%
\url{https://www.gelsight.com/}
\showURL{%
\tempurl}
\newblock
\shownote{Accessed: 2026-04-28}.


\bibitem[{GelSight}(2026b)]%
        {gelsightMiniDatasheet}
\bibfield{author}{\bibinfo{person}{{GelSight}}.}
  \bibinfo{year}{2026}\natexlab{b}.
\newblock \bibinfo{title}{{GelSight Mini Datasheet}}.
\newblock
\urldef\tempurl%
\url{https://www.gelsight.com/wp-content/uploads/productsheet/Mini/GelSight_Datasheet_GSMini.pdf}
\showURL{%
\tempurl}
\newblock
\shownote{Accessed: 2026-04-28}.


\bibitem[{GelSight}(2026c)]%
        {gelsightMiniProduct}
\bibfield{author}{\bibinfo{person}{{GelSight}}.}
  \bibinfo{year}{2026}\natexlab{c}.
\newblock \bibinfo{title}{{GelSight Mini Tactile Sensor}}.
\newblock
\urldef\tempurl%
\url{https://www.gelsight.com/gelsightmini/}
\showURL{%
\tempurl}
\newblock
\shownote{Accessed: 2026-04-28}.


\bibitem[Hu et~al\mbox{.}(2024)]%
        {hu2024magneticskin}
\bibfield{author}{\bibinfo{person}{Hao Hu}, \bibinfo{person}{Chengqian Zhang},
  \bibinfo{person}{Xinyi Lai}, \bibinfo{person}{Huangzhe Dai},
  \bibinfo{person}{Chengfeng Pan}, \bibinfo{person}{Haonan Sun},
  \bibinfo{person}{Daofan Tang}, \bibinfo{person}{Zhezai Hu},
  \bibinfo{person}{Jianzhong Fu}, \bibinfo{person}{Tiefeng Li}, {and}
  \bibinfo{person}{Peng Zhao}.} \bibinfo{year}{2024}\natexlab{}.
\newblock \showarticletitle{Large-Area Magnetic Skin for Multi-Point and
  Multi-Scale Tactile Sensing with Super-Resolution}.
\newblock \bibinfo{journal}{\emph{npj Flexible Electronics}}
  \bibinfo{volume}{8}, Article \bibinfo{articleno}{42} (\bibinfo{year}{2024}).
\newblock
\href{https://doi.org/10.1038/s41528-024-00325-z}{doi:\nolinkurl{10.1038/s41528-024-00325-z}}


\bibitem[Ji et~al\mbox{.}(2021)]%
        {ji2021poltergeist}
\bibfield{author}{\bibinfo{person}{Xiaoyu Ji}, \bibinfo{person}{Yushi Cheng},
  \bibinfo{person}{Yuepeng Zhang}, \bibinfo{person}{Kai Wang},
  \bibinfo{person}{Chen Yan}, \bibinfo{person}{Wenyuan Xu}, {and}
  \bibinfo{person}{Kevin Fu}.} \bibinfo{year}{2021}\natexlab{}.
\newblock \showarticletitle{Poltergeist: Acoustic Adversarial Machine Learning
  against Cameras and Computer Vision}. In \bibinfo{booktitle}{\emph{42nd
  {IEEE} Symposium on Security and Privacy, {SP} 2021, San Francisco, CA, USA,
  24-27 May 2021}}. \bibinfo{publisher}{{IEEE}}, \bibinfo{pages}{160--175}.
\newblock
\href{https://doi.org/10.1109/SP40001.2021.00091}{doi:\nolinkurl{10.1109/SP40001.2021.00091}}


\bibitem[Khamis et~al\mbox{.}(2021)]%
        {khamis2021friction}
\bibfield{author}{\bibinfo{person}{Heba Khamis}, \bibinfo{person}{Benjamin
  Xia}, {and} \bibinfo{person}{Stephen~J. Redmond}.}
  \bibinfo{year}{2021}\natexlab{}.
\newblock \showarticletitle{Real-Time Friction Estimation for Grip Force
  Control}. In \bibinfo{booktitle}{\emph{Proceedings of the 2021 IEEE
  International Conference on Robotics and Automation}}.
  \bibinfo{pages}{1608--1614}.
\newblock
\href{https://doi.org/10.1109/ICRA48506.2021.9561640}{doi:\nolinkurl{10.1109/ICRA48506.2021.9561640}}


\bibitem[Kulecki et~al\mbox{.}(2021)]%
        {kulecki2021practical}
\bibfield{author}{\bibinfo{person}{Bart{\l}omiej Kulecki},
  \bibinfo{person}{Kamil M{\l}odzikowski}, \bibinfo{person}{Rafa{\l} Staszak},
  {and} \bibinfo{person}{Dominik Belter}.} \bibinfo{year}{2021}\natexlab{}.
\newblock \showarticletitle{Practical Aspects of Detection and Grasping Objects
  by a Mobile Manipulating Robot}.
\newblock \bibinfo{journal}{\emph{Industrial Robot}} \bibinfo{volume}{48},
  \bibinfo{number}{5} (\bibinfo{year}{2021}), \bibinfo{pages}{688--699}.
\newblock
\href{https://doi.org/10.1108/IR-10-2020-0242}{doi:\nolinkurl{10.1108/IR-10-2020-0242}}


\bibitem[Kune et~al\mbox{.}(2013)]%
        {kune2013ghost}
\bibfield{author}{\bibinfo{person}{Denis~Foo Kune}, \bibinfo{person}{John~D.
  Backes}, \bibinfo{person}{Shane~S. Clark}, \bibinfo{person}{Daniel~B.
  Kramer}, \bibinfo{person}{Matthew~R. Reynolds}, \bibinfo{person}{Kevin Fu},
  \bibinfo{person}{Yongdae Kim}, {and} \bibinfo{person}{Wenyuan Xu}.}
  \bibinfo{year}{2013}\natexlab{}.
\newblock \showarticletitle{Ghost Talk: Mitigating {EMI} Signal Injection
  Attacks against Analog Sensors}. In \bibinfo{booktitle}{\emph{2013 {IEEE}
  Symposium on Security and Privacy, {SP} 2013, Berkeley, CA, USA, May 19-22,
  2013}}. \bibinfo{publisher}{{IEEE} Computer Society},
  \bibinfo{pages}{145--159}.
\newblock
\href{https://doi.org/10.1109/SP.2013.20}{doi:\nolinkurl{10.1109/SP.2013.20}}


\bibitem[Lambeta et~al\mbox{.}(2020)]%
        {lambeta2020digit}
\bibfield{author}{\bibinfo{person}{Mike Lambeta}, \bibinfo{person}{Po{-}Wei
  Chou}, \bibinfo{person}{Stephen Tian}, \bibinfo{person}{Brian~H. Yang},
  \bibinfo{person}{Benjamin Maloon}, \bibinfo{person}{Victoria~Rose Most},
  \bibinfo{person}{Dave Stroud}, \bibinfo{person}{Raymond Santos},
  \bibinfo{person}{Ahmad Byagowi}, \bibinfo{person}{Gregg Kammerer},
  \bibinfo{person}{Dinesh Jayaraman}, {and} \bibinfo{person}{Roberto
  Calandra}.} \bibinfo{year}{2020}\natexlab{}.
\newblock \bibinfo{title}{{DIGIT:} {A} Novel Design for a Low-Cost Compact
  High-Resolution Tactile Sensor with Application to In-Hand Manipulation}.
\newblock
\showeprint[arXiv]{2005.14679}
\urldef\tempurl%
\url{https://arxiv.org/abs/2005.14679}
\showURL{%
\tempurl}


\bibitem[Le~Signor et~al\mbox{.}(2023)]%
        {LeSignor2023MassManufacturable}
\bibfield{author}{\bibinfo{person}{Th{\'e}o Le~Signor},
  \bibinfo{person}{Nicolas Dupr{\'e}}, \bibinfo{person}{Jeroen Didden},
  \bibinfo{person}{Eugene Lomakin}, {and} \bibinfo{person}{Ga{\"e}l Close}.}
  \bibinfo{year}{2023}\natexlab{}.
\newblock \showarticletitle{Mass-Manufacturable 3D Magnetic Force Sensor for
  Robotic Grasping and Slip Detection}.
\newblock \bibinfo{journal}{\emph{Sensors}} \bibinfo{volume}{23},
  \bibinfo{number}{6}, Article \bibinfo{articleno}{3031}
  (\bibinfo{year}{2023}), \bibinfo{numpages}{18}~pages.
\newblock
\href{https://doi.org/10.3390/s23063031}{doi:\nolinkurl{10.3390/s23063031}}


\bibitem[Li et~al\mbox{.}(2020)]%
        {li2020tactile}
\bibfield{author}{\bibinfo{person}{Qiang Li}, \bibinfo{person}{Oliver Kroemer},
  \bibinfo{person}{Zhe Su}, \bibinfo{person}{Filipe~Fernandes Veiga},
  \bibinfo{person}{Mohsen Kaboli}, {and} \bibinfo{person}{Helge~Joachim
  Ritter}.} \bibinfo{year}{2020}\natexlab{}.
\newblock \showarticletitle{A Review of Tactile Information: Perception and
  Action Through Touch}.
\newblock \bibinfo{journal}{\emph{IEEE Transactions on Robotics}}
  \bibinfo{volume}{36}, \bibinfo{number}{6} (\bibinfo{year}{2020}),
  \bibinfo{pages}{1619--1634}.
\newblock
\href{https://doi.org/10.1109/TRO.2020.3003230}{doi:\nolinkurl{10.1109/TRO.2020.3003230}}


\bibitem[Luo et~al\mbox{.}(2024)]%
        {luo2024robotics}
\bibfield{author}{\bibinfo{person}{Jing Luo}, \bibinfo{person}{Xiangyu Zhou},
  \bibinfo{person}{Chao Zeng}, \bibinfo{person}{Yiming Jiang},
  \bibinfo{person}{Wen Qi}, \bibinfo{person}{Kui Xiang}, \bibinfo{person}{Muye
  Pang}, {and} \bibinfo{person}{Biwei Tang}.} \bibinfo{year}{2024}\natexlab{}.
\newblock \showarticletitle{Robotics Perception and Control: Key Technologies
  and Applications}.
\newblock \bibinfo{journal}{\emph{Micromachines}} \bibinfo{volume}{15},
  \bibinfo{number}{4}, Article \bibinfo{articleno}{531} (\bibinfo{year}{2024}).
\newblock
\href{https://doi.org/10.3390/mi15040531}{doi:\nolinkurl{10.3390/mi15040531}}


\bibitem[Man et~al\mbox{.}(2024)]%
        {man2024bionic}
\bibfield{author}{\bibinfo{person}{Jiandong Man}, \bibinfo{person}{Zhenhu Jin},
  {and} \bibinfo{person}{Jiamin Chen}.} \bibinfo{year}{2024}\natexlab{}.
\newblock \showarticletitle{Magnetic Tactile Sensor with Bionic Hair Array for
  Sliding Sensing and Object Recognition}.
\newblock \bibinfo{journal}{\emph{Advanced Science}} \bibinfo{volume}{11},
  \bibinfo{number}{12}, Article \bibinfo{articleno}{2306832}
  (\bibinfo{year}{2024}).
\newblock
\href{https://doi.org/10.1002/advs.202306832}{doi:\nolinkurl{10.1002/advs.202306832}}


\bibitem[Mao et~al\mbox{.}(2024)]%
        {mao2024multimodal}
\bibfield{author}{\bibinfo{person}{Qian Mao}, \bibinfo{person}{Zijian Liao},
  \bibinfo{person}{Jinfeng Yuan}, {and} \bibinfo{person}{Rong Zhu}.}
  \bibinfo{year}{2024}\natexlab{}.
\newblock \showarticletitle{Multimodal Tactile Sensing Fused with Vision for
  Dexterous Robotic Housekeeping}.
\newblock \bibinfo{journal}{\emph{Nature Communications}}
  \bibinfo{volume}{15}, Article \bibinfo{articleno}{6871}
  (\bibinfo{year}{2024}).
\newblock
\href{https://doi.org/10.1038/s41467-024-51261-5}{doi:\nolinkurl{10.1038/s41467-024-51261-5}}


\bibitem[Melexis(2025)]%
        {melexis_mlx90393_datasheet}
Melexis \bibinfo{year}{2025}\natexlab{}.
\newblock \bibinfo{booktitle}{\emph{{MLX90393
  Triaxis\textsuperscript{\textregistered} Micropower Magnetometer
  Datasheet}}}.
\newblock Melexis.
\newblock
\urldef\tempurl%
\url{https://www.melexis.com/-/media/files/documents/datasheets/mlx90393-datasheet-melexis.pdf}
\showURL{%
\tempurl}
\newblock
\shownote{Revision 012, November 2025}.


\bibitem[Meribout et~al\mbox{.}(2024)]%
        {meribout2024tactile}
\bibfield{author}{\bibinfo{person}{Mahmoud Meribout}, \bibinfo{person}{Natnael
  {Abule Takele}}, \bibinfo{person}{Olyad Derege}, \bibinfo{person}{Nidal
  Rifiki}, \bibinfo{person}{Mohamed {El Khalil}}, \bibinfo{person}{Varun
  Tiwari}, {and} \bibinfo{person}{Jing Zhong}.}
  \bibinfo{year}{2024}\natexlab{}.
\newblock \showarticletitle{Tactile Sensors: A Review}.
\newblock \bibinfo{journal}{\emph{Measurement}}  \bibinfo{volume}{238}, Article
  \bibinfo{articleno}{115332} (\bibinfo{year}{2024}).
\newblock
\href{https://doi.org/10.1016/j.measurement.2024.115332}{doi:\nolinkurl{10.1016/j.measurement.2024.115332}}


\bibitem[Oh et~al\mbox{.}(2017)]%
        {oh2017magnetic}
\bibfield{author}{\bibinfo{person}{Sunjong Oh}, \bibinfo{person}{Youngdo Jung},
  \bibinfo{person}{Seonggi Kim}, \bibinfo{person}{SungJoon Kim},
  \bibinfo{person}{Xinghao Hu}, \bibinfo{person}{Hyuneui Lim}, {and}
  \bibinfo{person}{CheolGi Kim}.} \bibinfo{year}{2017}\natexlab{}.
\newblock \showarticletitle{Remote Tactile Sensing System Integrated with
  Magnetic Synapse}.
\newblock \bibinfo{journal}{\emph{Scientific Reports}}  \bibinfo{volume}{7},
  Article \bibinfo{articleno}{16963} (\bibinfo{year}{2017}).
\newblock
\href{https://doi.org/10.1038/s41598-017-17277-2}{doi:\nolinkurl{10.1038/s41598-017-17277-2}}


\bibitem[Pattabiraman et~al\mbox{.}(2025)]%
        {pattabiraman2025eflesh}
\bibfield{author}{\bibinfo{person}{Venkatesh Pattabiraman},
  \bibinfo{person}{Zizhou Huang}, \bibinfo{person}{Daniele Panozzo},
  \bibinfo{person}{Denis Zorin}, \bibinfo{person}{Lerrel Pinto}, {and}
  \bibinfo{person}{Raunaq Bhirangi}.} \bibinfo{year}{2025}\natexlab{}.
\newblock \bibinfo{title}{{eFlesh}: Highly Customizable Magnetic Touch Sensing
  Using Cut-Cell Microstructures}.
\newblock
\showeprint[arxiv]{2506.09994}
\href{https://doi.org/10.48550/arXiv.2506.09994}{doi:\nolinkurl{10.48550/arXiv.2506.09994}}


\bibitem[{PaXini}(2026)]%
        {paxini2026px6axgen3}
\bibfield{author}{\bibinfo{person}{{PaXini}}.} \bibinfo{year}{2026}\natexlab{}.
\newblock \bibinfo{title}{{PX-6AX GEN3: Multi-dimensional Tactile Sensor}}.
\newblock
\urldef\tempurl%
\url{https://www.paxini.com/cn/ax/gen3}
\showURL{%
\tempurl}
\newblock
\shownote{Accessed: 2026-04-28}.


\bibitem[{PaXini Tech}(2026a)]%
        {paxini2026itpu}
\bibfield{author}{\bibinfo{person}{{PaXini Tech}}.}
  \bibinfo{year}{2026}\natexlab{a}.
\newblock \bibinfo{title}{{ITPU technology integrates multi-dimensional tactile
  sensing}}.
\newblock
\urldef\tempurl%
\url{https://www.paxini.com/cn/itpu}
\showURL{%
\tempurl}
\newblock
\shownote{Accessed: 2026-04-28}.


\bibitem[{PaXini Tech}(2026b)]%
        {paxini2026official}
\bibfield{author}{\bibinfo{person}{{PaXini Tech}}.}
  \bibinfo{year}{2026}\natexlab{b}.
\newblock \bibinfo{title}{{PaXini official}}.
\newblock
\urldef\tempurl%
\url{https://www.paxini.com/cn/}
\showURL{%
\tempurl}
\newblock
\shownote{Accessed: 2026-04-28}.


\bibitem[{PaXini Tech}(2026c)]%
        {paxini2026px6axgen3mall}
\bibfield{author}{\bibinfo{person}{{PaXini Tech}}.}
  \bibinfo{year}{2026}\natexlab{c}.
\newblock \bibinfo{title}{{PX-6AX GEN3: PaXini Mall Product Page}}.
\newblock
\urldef\tempurl%
\url{https://mall.paxini.com/product/itpu/6894488c74e45ce616705b80}
\showURL{%
\tempurl}
\newblock
\shownote{Accessed: 2026-04-28}.


\bibitem[{Robotiq}(2019)]%
        {robotiq2fManual}
\bibfield{author}{\bibinfo{person}{{Robotiq}}.}
  \bibinfo{year}{2019}\natexlab{}.
\newblock \bibinfo{title}{{2F-85 and 2F-140 Adaptive Grippers Instruction
  Manual}}.
\newblock
\urldef\tempurl%
\url{https://assets.robotiq.com/website-assets/support_documents/document/online/2F-85_2F-140_TM_InstructionManual_HTML5_20190503.zip/2F-85_2F-140_TM_InstructionManual_HTML5/Content/4.%20Control.htm}
\showURL{%
\tempurl}
\newblock
\shownote{Accessed: 2026-04-30}.


\bibitem[Schmitz et~al\mbox{.}(2011)]%
        {schmitz2011methods}
\bibfield{author}{\bibinfo{person}{Alexander Schmitz}, \bibinfo{person}{Perla
  Maiolino}, \bibinfo{person}{Marco Maggiali}, \bibinfo{person}{Lorenzo
  Natale}, \bibinfo{person}{Giorgio Cannata}, {and} \bibinfo{person}{Giorgio
  Metta}.} \bibinfo{year}{2011}\natexlab{}.
\newblock \showarticletitle{Methods and Technologies for the Implementation of
  Large-Scale Robot Tactile Sensors}.
\newblock \bibinfo{journal}{\emph{{IEEE} Trans. Robotics}}
  \bibinfo{volume}{27}, \bibinfo{number}{3} (\bibinfo{year}{2011}),
  \bibinfo{pages}{389--400}.
\newblock
\href{https://doi.org/10.1109/TRO.2011.2132930}{doi:\nolinkurl{10.1109/TRO.2011.2132930}}


\bibitem[Signor et~al\mbox{.}(2022)]%
        {le2022gradiometric}
\bibfield{author}{\bibinfo{person}{Th{\'{e}}o~Le Signor},
  \bibinfo{person}{Nicolas Dupr{\'{e}}}, {and} \bibinfo{person}{Ga{\"{e}}l
  Close}.} \bibinfo{year}{2022}\natexlab{}.
\newblock \showarticletitle{A Gradiometric Magnetic Force Sensor Immune to
  Stray Magnetic Fields for Robotic Hands and Grippers}.
\newblock \bibinfo{journal}{\emph{{IEEE} Robotics Autom. Lett.}}
  \bibinfo{volume}{7}, \bibinfo{number}{2} (\bibinfo{year}{2022}),
  \bibinfo{pages}{3070--3076}.
\newblock
\href{https://doi.org/10.1109/LRA.2022.3146507}{doi:\nolinkurl{10.1109/LRA.2022.3146507}}


\bibitem[Silva et~al\mbox{.}(2019)]%
        {silva2019mass}
\bibfield{author}{\bibinfo{person}{Andr{\'e} Silva}, \bibinfo{person}{Maria
  Brites}, \bibinfo{person}{Tiago Paulino}, {and} \bibinfo{person}{Plinio
  Moreno}.} \bibinfo{year}{2019}\natexlab{}.
\newblock \showarticletitle{Estimation of Lightweight Object's Mass by a
  Humanoid Robot During a Precision Grip with Soft Tactile Sensors}. In
  \bibinfo{booktitle}{\emph{Proceedings of the 2019 Third IEEE International
  Conference on Robotic Computing}}. \bibinfo{pages}{344--348}.
\newblock
\href{https://doi.org/10.1109/IRC.2019.00062}{doi:\nolinkurl{10.1109/IRC.2019.00062}}


\bibitem[{SingleTact}(2026a)]%
        {singletact2026calibration}
\bibfield{author}{\bibinfo{person}{{SingleTact}}.}
  \bibinfo{year}{2026}\natexlab{a}.
\newblock \bibinfo{title}{{Calibrated vs Uncalibrated Force Sensor}}.
\newblock
\urldef\tempurl%
\url{https://www.singletact.com/calibration}
\showURL{%
\tempurl}
\newblock
\shownote{Accessed: 2026-04-28}.


\bibitem[{SingleTact}(2026b)]%
        {singletact2026official}
\bibfield{author}{\bibinfo{person}{{SingleTact}}.}
  \bibinfo{year}{2026}\natexlab{b}.
\newblock \bibinfo{title}{{SingleTact Micro Force Sensors}}.
\newblock
\urldef\tempurl%
\url{https://www.singletact.com/}
\showURL{%
\tempurl}
\newblock
\shownote{Accessed: 2026-04-28}.


\bibitem[Stassi et~al\mbox{.}(2014)]%
        {stassi2014flexible}
\bibfield{author}{\bibinfo{person}{Stefano Stassi}, \bibinfo{person}{Valentina
  Cauda}, \bibinfo{person}{Giancarlo Canavese}, {and}
  \bibinfo{person}{Candido~Fabrizio Pirri}.} \bibinfo{year}{2014}\natexlab{}.
\newblock \showarticletitle{Flexible Tactile Sensing Based on Piezoresistive
  Composites: {A} Review}.
\newblock \bibinfo{journal}{\emph{Sensors}} \bibinfo{volume}{14},
  \bibinfo{number}{3} (\bibinfo{year}{2014}), \bibinfo{pages}{5296--5332}.
\newblock
\href{https://doi.org/10.3390/S140305296}{doi:\nolinkurl{10.3390/S140305296}}


\bibitem[{Tekscan}(2026)]%
        {tekscan2026official}
\bibfield{author}{\bibinfo{person}{{Tekscan}}.}
  \bibinfo{year}{2026}\natexlab{}.
\newblock \bibinfo{title}{{Pressure Mapping, Force Measurement, and Tactile
  Sensors}}.
\newblock
\urldef\tempurl%
\url{https://www.tekscan.com/}
\showURL{%
\tempurl}
\newblock
\shownote{Accessed: 2026-04-28}.


\bibitem[Tomo et~al\mbox{.}(2016)]%
        {tomo2016hall}
\bibfield{author}{\bibinfo{person}{Tito~Pradhono Tomo}, \bibinfo{person}{Sophon
  Somlor}, \bibinfo{person}{Alexander Schmitz}, \bibinfo{person}{Lorenzo
  Jamone}, \bibinfo{person}{Weijie Huang}, \bibinfo{person}{Harris Kristanto},
  {and} \bibinfo{person}{Shigeki Sugano}.} \bibinfo{year}{2016}\natexlab{}.
\newblock \showarticletitle{Design and Characterization of a Three-Axis Hall
  Effect-Based Soft Skin Sensor}.
\newblock \bibinfo{journal}{\emph{Sensors}} \bibinfo{volume}{16},
  \bibinfo{number}{4}, Article \bibinfo{articleno}{491} (\bibinfo{year}{2016}).
\newblock
\href{https://doi.org/10.3390/s16040491}{doi:\nolinkurl{10.3390/s16040491}}


\bibitem[Trippel et~al\mbox{.}(2017)]%
        {trippel2017walnut}
\bibfield{author}{\bibinfo{person}{Timothy Trippel}, \bibinfo{person}{Ofir
  Weisse}, \bibinfo{person}{Wenyuan Xu}, \bibinfo{person}{Peter Honeyman},
  {and} \bibinfo{person}{Kevin Fu}.} \bibinfo{year}{2017}\natexlab{}.
\newblock \showarticletitle{{WALNUT:} Waging Doubt on the Integrity of {MEMS}
  Accelerometers with Acoustic Injection Attacks}. In
  \bibinfo{booktitle}{\emph{2017 {IEEE} European Symposium on Security and
  Privacy, EuroS{\&}P 2017, Paris, France, April 26-28, 2017}}.
  \bibinfo{publisher}{{IEEE}}, \bibinfo{pages}{3--18}.
\newblock
\href{https://doi.org/10.1109/EUROSP.2017.42}{doi:\nolinkurl{10.1109/EUROSP.2017.42}}


\bibitem[{Unitree Robotics}(2026)]%
        {unitree_g1_developer}
\bibfield{author}{\bibinfo{person}{{Unitree Robotics}}.}
  \bibinfo{year}{2026}\natexlab{}.
\newblock \bibinfo{title}{{G1 Developer Documentation}}.
\newblock
\urldef\tempurl%
\url{https://support.unitree.com/home/zh/G1_developer/about_G1}
\showURL{%
\tempurl}
\newblock
\shownote{Accessed: 2026-04-30}.


\bibitem[{Universal Robots}(2026a)]%
        {urRTDEGuide}
\bibfield{author}{\bibinfo{person}{{Universal Robots}}.}
  \bibinfo{year}{2026}\natexlab{a}.
\newblock \bibinfo{title}{Real-Time Data Exchange ({RTDE}) Guide}.
\newblock
\urldef\tempurl%
\url{https://docs.universal-robots.com/tutorials/communication-protocol-tutorials/rtde-guide.html}
\showURL{%
\tempurl}
\newblock
\shownote{Accessed: 2026-04-30}.


\bibitem[{Universal Robots}(2026b)]%
        {ur5eSpec}
\bibfield{author}{\bibinfo{person}{{Universal Robots}}.}
  \bibinfo{year}{2026}\natexlab{b}.
\newblock \bibinfo{title}{{Technical Specifications UR5e}}.
\newblock
\urldef\tempurl%
\url{https://www.universal-robots.com/manuals/EN/HTML/SW10_6/Content/prod-usr-man/hardware/arm_e-Series/UR5e/H_g5_sections/appendix_g5/tech_spec_sheet.htm}
\showURL{%
\tempurl}
\newblock
\shownote{Accessed: 2026-04-30}.


\bibitem[Ward-Cherrier et~al\mbox{.}(2018)]%
        {ward2018tactip}
\bibfield{author}{\bibinfo{person}{Benjamin Ward-Cherrier},
  \bibinfo{person}{Nicholas Pestell}, \bibinfo{person}{Luke Cramphorn},
  \bibinfo{person}{Benjamin Winstone}, \bibinfo{person}{Maria~Elena
  Giannaccini}, \bibinfo{person}{Jonathan Rossiter}, {and}
  \bibinfo{person}{Nathan~F Lepora}.} \bibinfo{year}{2018}\natexlab{}.
\newblock \showarticletitle{The tactip family: Soft optical tactile sensors
  with 3d-printed biomimetic morphologies}.
\newblock \bibinfo{journal}{\emph{Soft robotics}} \bibinfo{volume}{5},
  \bibinfo{number}{2} (\bibinfo{year}{2018}), \bibinfo{pages}{216--227}.
\newblock
\href{https://doi.org/10.1089/soro.2017.0052}{doi:\nolinkurl{10.1089/soro.2017.0052}}


\bibitem[{XELA Robotics}(2026a)]%
        {xela2026official}
\bibfield{author}{\bibinfo{person}{{XELA Robotics}}.}
  \bibinfo{year}{2026}\natexlab{a}.
\newblock \bibinfo{title}{{Physical AI with a Human Touch}}.
\newblock
\urldef\tempurl%
\url{https://xelarobotics.com/}
\showURL{%
\tempurl}
\newblock
\shownote{Accessed: 2026-04-28}.


\bibitem[{XELA Robotics}(2026b)]%
        {xela2026technology}
\bibfield{author}{\bibinfo{person}{{XELA Robotics}}.}
  \bibinfo{year}{2026}\natexlab{b}.
\newblock \bibinfo{title}{{Tactile Sensing Technology for Robots}}.
\newblock
\urldef\tempurl%
\url{https://xelarobotics.com/technology/}
\showURL{%
\tempurl}
\newblock
\shownote{Accessed: 2026-04-28}.


\bibitem[Xu et~al\mbox{.}(2018)]%
        {xu2018analyzing}
\bibfield{author}{\bibinfo{person}{Wenyuan Xu}, \bibinfo{person}{Chen Yan},
  \bibinfo{person}{Weibin Jia}, \bibinfo{person}{Xiaoyu Ji}, {and}
  \bibinfo{person}{Jianhao Liu}.} \bibinfo{year}{2018}\natexlab{}.
\newblock \showarticletitle{Analyzing and Enhancing the Security of Ultrasonic
  Sensors for Autonomous Vehicles}.
\newblock \bibinfo{journal}{\emph{{IEEE} Internet Things J.}}
  \bibinfo{volume}{5}, \bibinfo{number}{6} (\bibinfo{year}{2018}),
  \bibinfo{pages}{5015--5029}.
\newblock
\href{https://doi.org/10.1109/JIOT.2018.2867917}{doi:\nolinkurl{10.1109/JIOT.2018.2867917}}


\bibitem[Xu and She(2024)]%
        {xu2024letacmpc}
\bibfield{author}{\bibinfo{person}{Zhengtong Xu} {and} \bibinfo{person}{Yu
  She}.} \bibinfo{year}{2024}\natexlab{}.
\newblock \showarticletitle{{LeTac-MPC}: Learning Model Predictive Control for
  Tactile-Reactive Grasping}.
\newblock \bibinfo{journal}{\emph{IEEE Transactions on Robotics}}
  \bibinfo{volume}{40} (\bibinfo{year}{2024}), \bibinfo{pages}{4376--4395}.
\newblock
\href{https://doi.org/10.1109/TRO.2024.3463470}{doi:\nolinkurl{10.1109/TRO.2024.3463470}}


\bibitem[{Yahboom}(2026)]%
        {yahboom_dofbot_pi}
\bibfield{author}{\bibinfo{person}{{Yahboom}}.}
  \bibinfo{year}{2026}\natexlab{}.
\newblock \bibinfo{title}{{DOFBOT-Pi}}.
\newblock
\urldef\tempurl%
\url{https://www.yahboom.net/study/Dofbot-Pi}
\showURL{%
\tempurl}
\newblock
\shownote{Accessed: 2026-04-30}.


\bibitem[Yan et~al\mbox{.}(2021b)]%
        {yan2019feasibility}
\bibfield{author}{\bibinfo{person}{Chen Yan}, \bibinfo{person}{Guoming Zhang},
  \bibinfo{person}{Xiaoyu Ji}, \bibinfo{person}{Tianchen Zhang},
  \bibinfo{person}{Taimin Zhang}, {and} \bibinfo{person}{Wenyuan Xu}.}
  \bibinfo{year}{2021}\natexlab{b}.
\newblock \showarticletitle{The Feasibility of Injecting Inaudible Voice
  Commands to Voice Assistants}.
\newblock \bibinfo{journal}{\emph{{IEEE} Trans. Dependable Secur. Comput.}}
  \bibinfo{volume}{18}, \bibinfo{number}{3} (\bibinfo{year}{2021}),
  \bibinfo{pages}{1108--1124}.
\newblock
\href{https://doi.org/10.1109/TDSC.2019.2906165}{doi:\nolinkurl{10.1109/TDSC.2019.2906165}}


\bibitem[Yan et~al\mbox{.}(2021a)]%
        {yan2021softmagnetic}
\bibfield{author}{\bibinfo{person}{Youcan Yan}, \bibinfo{person}{Zhe Hu},
  \bibinfo{person}{Zhengbao Yang}, \bibinfo{person}{Wenzhen Yuan},
  \bibinfo{person}{Chaoyang Song}, \bibinfo{person}{Jia Pan}, {and}
  \bibinfo{person}{Yajing Shen}.} \bibinfo{year}{2021}\natexlab{a}.
\newblock \showarticletitle{Soft Magnetic Skin for Super-Resolution Tactile
  Sensing with Force Self-Decoupling}.
\newblock \bibinfo{journal}{\emph{Science Robotics}} \bibinfo{volume}{6},
  \bibinfo{number}{51}, Article \bibinfo{articleno}{eabc8801}
  (\bibinfo{year}{2021}).
\newblock
\href{https://doi.org/10.1126/scirobotics.abc8801}{doi:\nolinkurl{10.1126/scirobotics.abc8801}}


\bibitem[Yuan et~al\mbox{.}(2017a)]%
        {yuan2017gelsight}
\bibfield{author}{\bibinfo{person}{Wenzhen Yuan}, \bibinfo{person}{Siyuan
  Dong}, {and} \bibinfo{person}{Edward~H. Adelson}.}
  \bibinfo{year}{2017}\natexlab{a}.
\newblock \showarticletitle{GelSight: High-Resolution Robot Tactile Sensors for
  Estimating Geometry and Force}.
\newblock \bibinfo{journal}{\emph{Sensors}} \bibinfo{volume}{17},
  \bibinfo{number}{12} (\bibinfo{year}{2017}), \bibinfo{pages}{2762}.
\newblock
\href{https://doi.org/10.3390/S17122762}{doi:\nolinkurl{10.3390/S17122762}}


\bibitem[Yuan et~al\mbox{.}(2017b)]%
        {yuan2017hardness}
\bibfield{author}{\bibinfo{person}{Wenzhen Yuan}, \bibinfo{person}{Chenzhuo
  Zhu}, \bibinfo{person}{Andrew Owens}, \bibinfo{person}{Mandayam~A.
  Srinivasan}, {and} \bibinfo{person}{Edward~H. Adelson}.}
  \bibinfo{year}{2017}\natexlab{b}.
\newblock \showarticletitle{Shape-Independent Hardness Estimation Using Deep
  Learning and a {GelSight} Tactile Sensor}. In
  \bibinfo{booktitle}{\emph{Proceedings of the 2017 IEEE International
  Conference on Robotics and Automation}}. \bibinfo{pages}{951--958}.
\newblock
\href{https://doi.org/10.1109/ICRA.2017.7989116}{doi:\nolinkurl{10.1109/ICRA.2017.7989116}}


\bibitem[Zhang et~al\mbox{.}(2017)]%
        {zhang2017dolphinattack}
\bibfield{author}{\bibinfo{person}{Guoming Zhang}, \bibinfo{person}{Chen Yan},
  \bibinfo{person}{Xiaoyu Ji}, \bibinfo{person}{Tianchen Zhang},
  \bibinfo{person}{Taimin Zhang}, {and} \bibinfo{person}{Wenyuan Xu}.}
  \bibinfo{year}{2017}\natexlab{}.
\newblock \showarticletitle{DolphinAttack: Inaudible Voice Commands}. In
  \bibinfo{booktitle}{\emph{Proceedings of the 2017 {ACM} {SIGSAC} Conference
  on Computer and Communications Security, {CCS} 2017, Dallas, TX, USA, October
  30 - November 03, 2017}}, \bibfield{editor}{\bibinfo{person}{Bhavani
  Thuraisingham}, \bibinfo{person}{David Evans}, \bibinfo{person}{Tal Malkin},
  {and} \bibinfo{person}{Dongyan Xu}} (Eds.). \bibinfo{publisher}{{ACM}},
  \bibinfo{pages}{103--117}.
\newblock
\href{https://doi.org/10.1145/3133956.3134052}{doi:\nolinkurl{10.1145/3133956.3134052}}


\bibitem[Zhao et~al\mbox{.}(2025)]%
        {zhao2025mpddm}
\bibfield{author}{\bibinfo{person}{Wenzheng Zhao}, \bibinfo{person}{Kruthika
  Gangaraju}, {and} \bibinfo{person}{Fengpei Yuan}.}
  \bibinfo{year}{2025}\natexlab{}.
\newblock \showarticletitle{Multimodal Perception-Driven Decision-Making for
  Human-Robot Interaction: A Survey}.
\newblock \bibinfo{journal}{\emph{Frontiers in Robotics and AI}}
  \bibinfo{volume}{12}, Article \bibinfo{articleno}{1604472}
  (\bibinfo{year}{2025}).
\newblock
\href{https://doi.org/10.3389/frobt.2025.1604472}{doi:\nolinkurl{10.3389/frobt.2025.1604472}}


\end{thebibliography}
\appendix %

\section{Open Science} %
In the publicly released artifacts and demonstrations at:
\url{https://phantomtac.org/},
we provide the following materials:
\begin{itemize}
\item Python and MATLAB scripts for generating spatial heatmaps of three-dimensional force responses;
\item Python source code for RBF model fitting and blind testing;
\item Demonstration videos showing representative end-to-end attack outcomes on the Dofbot-pi4B robotic arm and Unitree G1 humanoid robot.
\end{itemize}

These materials support inspection of our modeling and visualization pipeline and provide qualitative evidence of the end-to-end experimental results. 
To limit direct misuse, we do not provide the complete model-specific attack-tag designs or detailed deployment parameters required for deterministic force injection.

\section{Ethical Considerations} 
This research demonstrates a physical-layer injection attack against magnetic tactile sensors. The attack can cause significant disruptions in robotic manipulation. 
These disruptions include abnormal gripping forces and object damage.

Recognizing the potential real-world implications of these vulnerabilities, we have strictly adhered to ethical guidelines throughout our research.
Specifically, all attack evaluations and system-level validations were conducted in a strictly controlled laboratory environment. Furthermore, safety measures, such as defining soft operational limits and emergency stop protocols, were active during testing to ensure the physical safety of the researchers. We responsibly disclosed the identified vulnerability to the affected sensor vendors.
By highlighting the potential threats in current single-modality tactile perception, we aim to raise awareness within the robotics community and advocate for the adoption of more robust, multi-modal defense mechanisms, such as the proprioceptive consistency checks proposed in Sec.~\ref{sec:Countermeasures}.

\end{document}